\documentclass[10pt]{article}

\usepackage{amsmath}
\usepackage{amssymb}

\usepackage{graphicx}

\usepackage{cite}

\usepackage{color} 

\usepackage[labelfont=bf,labelsep=period,justification=raggedright]{caption}

\makeatletter
\renewcommand{\@biblabel}[1]{\quad#1.}
\makeatother

\date{}

\usepackage{psfrag}
\usepackage{subfigure}

\begin{document}

% Title must be 150 characters or less
\begin{flushleft}
{\Large
\textbf{Computable Compressed Matrices}
}
% Insert Author names, affiliations and corresponding author email.
\\
Crysttian Arantes Paix\~{a}o$^{1\ast}$, 
Fl\'{a}vio Code\c{c}o Coelho$^{2}$, 
\\
\bf{1} Crysttian Arantes Paix\~{a}o Applied Mathematics School, Getulio Vargas Foundation, Rio de Janeiro, RJ, Brazil
\\
\bf{2} Fl\'{a}vio Code\c{c}o Coelho Applied Mathematics School, Getulio Vargas Foundation, Rio de Janeiro, RJ, Brazil
\\
$\ast$ E-mail: Corresponding crysttian.paixao@fgv.br
\end{flushleft}

% Please keep the abstract between 250 and 300 words
\section*{Abstract}
The biggest cost of
computing with large matrices in any modern computer is related to memory
latency and bandwidth. The average latency of modern RAM reads is 150 times
greater than a clock step of the processor\cite{alted2010modern}. Throughput is
a little better but still 25 times slower than the CPU can consume. The
application of bitstring compression allows for larger matrices to be moved
entirely to the cache memory of the computer, which has much better latency and
bandwidth (average latency of L1 cache is 3 to 4 clock steps). This allows for
massive performance gains as well as the ability to simulate much larger models
efficiently. In this work, we propose a methodology to compress matrices in such
a way that they retain their mathematical properties. Considerable compression
of the data is also achieved in the process Thus allowing for the computation of
much larger linear problems within the same memory constraints when compared
with the traditional representation of matrices.
% Please keep the Author Summary between 150 and 200 words
% Use first person. PLoS ONE authors please skip this step. 
% Author Summary not valid for PLoS ONE submissions.   
\section*{Author Summary}

\section*{Introduction}

Data compression is traditionally used to reduce storage resources usage and/or transmission costs\cite{salomon}. Compression techniques can be classified into lossy and lossless. Examples of lossy data compression are MP3 (audio), JPEG (image) and MPEG (video). In this paper we discuss the use of lossless compression for numerical data structures such as numerical arrays to achieve compression without losing the mathematical properties of the original data.  

Lossless compression methods usually exploit redundancies present in the data in order to find a shorter form of describing the same information content. For example, a dictionary-based compression, only stores the positions in which a given word occurs in a document, thus saving the space required to store all its repetitions\cite{salomon2}. 

Any kind of compression incurs some computational cost. Such costs often have to be paid twice since the data needs to be decompressed to be used for its original purpose. Sometimes computational costs are irrelevant, but the need to decompress for usage, can signify that the space saved with compression must be available when data is decompressed for usage, thus partially negating the advantages of compression.

Most if not all existing lossless compression methods were developed under the following usage paradigm: \textit{produce} $\rightarrow$ \textit{compress} $\rightarrow$ \textit{store} $\rightarrow$ \textit{uncompress} $\rightarrow$ \textit{use}. The focus of the present work is to allow a slightly different usage:  \textit{produce} $\rightarrow$ \textit{compress} $\rightarrow$ \textit{perform mathematical manipulations} and \textit{decompress} (only for human reading).

With the growth of data volumes and analytical demands, creative solutions are needed to efficiently store as well as consume it on demand. This issue is present in many areas of application, ranging from business to science\cite{lynch}, and is being called the \textit{Big Data} phenomenon. In the world of Big Data, the need to analyze data immediately after its coming into existence became the norm. And this analysis must take place, efficiently, within the confines of (RAM) memory. This kind of analyses are what is now known as streaming data analysis\cite{gaber2005mining}. Given a sufficiently dense stream of data, compression an decompression costs may become prohibitive. So having a way to compress data and keeping it compressed for the entire course of the analytical pipeline, is very desirable.

This paper will focus solely on numerical data which for the purpose of the applications is organized as   matrices. This is a most common data structure found in computational data analysis environments. The matrices compressed according to the methodologies proposed here should be able to undergo the same mathematical operations as the original uncompressed matrices, e.g. linear algebra manipulations. This way, the cost of compression is reduced to a single event of compression and no need of decompression except when displaying the results for human reading. The idea of operating with compressed arrays is relatively new\cite{yemliha2007compiler}, and it has yet to find mainstream applications to the field of numerical computations. One application which employs a form of compression is the sparse matrix linear algebra algorithms\cite{dodson1991sparse}, in this case there is no alteration in the standard encoding of the data, but only the non-zero elements of the matrices are stored and operated upon. 

Larger than RAM data structures can render traditional analytical algorithms impracticable. Currently, the technique most commonly used when dealing with large matrices for numerical computations, is memory mapping\cite{van2011numpy,big}. In memory mapping the matrix is allocated in a virtual contiguous address space which extends from memory into disk. Thus, larger than memory data structures can be manipulated as if they were in memory.  This technique has a big performance penalty due to lower access speeds of disk when compared to RAM. 

In this paper we present two methods for the lossless compression of (numerical) arrays. The methods involve the encoding of the numbers as strings of bits of variable length. The methods resemble the arithmetic coding\cite{bodden2007arithmetic} algorithm, but is cheaper to compute. We describe the process of compression and decompression, and study their efficiency under different applications. We also discuss the efficiency of the compression as a function of the distribution of the elements of the matrix. 

\section*{Methods}

\subsection*{Matrix compression}

To maintain mathematical equivalence with the original data for any arithmetic operations, we need to  maintain the structure of the matrix, i.e., the ability to acess any element given its row $i$ and column $j$ and also the numeric nature of its elements. In order to achieve compression we decided to exploit inefficiencies in the conventional way matrices are allocated in memory. The examples in this paper will be restricted to matrices with positive integer elements.

The compression method is as follows. Let $M_{r \times c}$ be a matrix, in which $r$ is the number of rows and $c$ the number of columns. Each element of this matrix, called $m_{ij}$, is a positive integer. In digital computers, all information is stored as binary code (base 2 numbers). However the conventional way to store arrays of integers is on a memory block sequence of fixed size (power of 2 numbers of bit), one for each element. The maximum size of a block is equal to the word size of the processor, which for most current CPUs is 64 bits. Some special number such as complex number may be encoded as two blocks instead of one. The size of the chunk of memory allocated to each number will determine their maximum  size (for integers) or their precision (for floating-point numbes). So for matrix $M$, the total memory allocated, assuming chunks of 64 bits, is given by $\mathcal{B} = r \times c \times 64$.

The number of bits allocated $\mathcal{B}$, is larger than the absolute minimum number of bits required to represent all the elements of $M$, since smaller integers, when converted to base 2, require less digits. From now on, when the numerical base will be explicitly notated when necessary to avoid confusion between binary and decimal integers.  

Let's consider an extreme example: a matrix composed exclusively of 0s and 1s (base 10). If the matrix type is set to 64-bit integers, 63 bits will be wasted per element of the matrix, since the minimum number of bits needed to store such a matrix is $\mathfrak{b} = r \times c \times 1$. The potential economy of bits $\xi$ can be represented by $\xi = \mathcal{B} - \mathfrak{b} =  r \times c \times 63$.

So it is evident that for any matrix whose greatest element requires less than 64 bits (or the fixed type of the matrix) to be represented, potential memory savings will grow linearly with the size of the matrix.

\subsubsection*{Method 1: The Supreme Minimum (SM)}

The SM method consists in determining the value of the greatest element of matrix $M$, which coincides with its supremum, $max M == \sup M$  and determine the minimum number of bits, $b(\sup M$)(Equation \ref{eq:01}), required to store it. We will use capital roman letters to denote uncompressed matrices and the corresponding lower case letter for the compressed version.

\begin{equation} \label{eq:01}
 b(\sup M) \approx \begin{cases}
	1, &  \text{ if } \sup M \in \{0,1\} \\ 
	\lfloor \log_2(\sup M)  \rfloor + 1,  & \text{ if } \sup M > 1  
	\end{cases}
\end{equation}

The allocation of memory still happens in the usual way, i.e., in fixed size 64-bit chunks, only that now, in the space required for a single 64 bit integer, we can store for example, an entire $8\times 8$ matrix of $0_{10}$ and $1_{10}$. 

Let's look at a concrete example: suppose that the greatest value to be stored in a matrix $M$ is $\max M=1023$. Therefore, the number of bits required to represent it is $10$ $(1111111111)$. Let the first 8 elements of $M$ be:

\begin{equation} \label{eq:02}
  M = \begin{bmatrix}
  900 & 1023 & 721 & 256 & 1 & 10 & 700 & 20 & \hdots\\ 
  \vdots & \vdots & \vdots & \vdots & \vdots & \vdots & \vdots & \vdots & \ddots 
  \end{bmatrix}
\end{equation}

These elements of $M$, in binary, are shown in Table \ref{tab:01}. It is evident that the number of bits required to represent any other element must be lower or equal to $10$. From now on the minimum number of bits required to represent a base 10 integer will be refered to as its bit-length.

\begin{table}[h]
 \centering
 \caption{Some elements of $M$ represented in binary base.}
 \begin{tabular}{cccc} 
  \hline 
  Element & Value  & Binary & Bit length\\
  \hline
  $M_{1,1}$ & 900  & 1110000100 & 10\\
  $M_{1,2}$ & 1023 & 1111111111 & 10\\
  $M_{1,3}$ & 721  & 1011010001 & 10\\
  $M_{1,4}$ & 256  & 100000000  & 9\\
  $M_{1,5}$ & 1    & 1          & 1\\
  $M_{1,6}$ & 10   & 1010       & 4\\
  $M_{1,7}$ & 700  & 1010111100 & 10\\
  $M_{1,8}$ & 20   & 10100      & 5\\
  \hline
 \end{tabular}
 \label{tab:01}
\end{table}

To store matrix $M$ it first has to be converted to base $2$ ($M_2$). Then it will be unraveled by column (column major, e.g. in Fortran) or by row (row major, e.g. in C) and its elements will be written as  fixed size adjacent chunks of memory. The size of each chunk is determined by the type associated with the matrix (typically 64 bits, but always a power of 2).

According to the SM method, having determined that each element will require at most 10 bits, we can divide the memory block corresponding to a single 64 bit integer into six 10-bit chunks which can each hold a single element of $M$. These 64-bit blocks will be called a bitstring. The remaining 4 bits will be used later. The number of bitstrings needed will be $\lfloor \frac{\dim(M) *b(\sup M)} {64} \rfloor +1$, where $dim(M)$ is the dimension of the matrix or its number of elements.

The final layout of the first 6 elements of $m$ in the first bitstring can be seen in \ref{eq:03}.

\begin{equation}\label{eq:03}
 bitstring_1 = 0000\underbrace{0000001010}_{10}\underbrace{0000000001}_{1}\underbrace{0100000000}_{256}\underbrace{1011010001}_{721}\underbrace{1111111111}_{1023}\underbrace{1110000100}_{900} 
\end{equation}

Here is a step-by-step description of the application of the SM method to matrix $M$:

\begin{enumerate}
 \item Element $M_{1,1}=900 = m_{1,1} = 1110000100$ is stored in the first 10-bit chunk of the element strip $bitstring[1]$ , which corresponds to bits 0 to 9 (read from right to left).
\begin{equation*}\label{eq:04}
bitstring_1 = 000000000000000000000000000000000000000000000000000000\underbrace{1110000100}_{900} 
\end{equation*}
 \item Element $M_{1,2}=1023$ is allocated in the second chunk, from bit 10 to bit 19.
\begin{equation*}\label{eq:05}
 bitstring_1 = 00000000000000000000000000000000000000000000\underbrace{1111111111}_{1023}\underbrace{1110000100}_{900} 
\end{equation*}
 \item Repeat for elements $M_{1,i}$ with $i=1,\ldots,6$ which are stored on the remaining chunks.
 \begin{equation*}\label{eq:06}
bitstring_1 = 0000\underbrace{0000001010}_{10}\underbrace{0000000001}_{1}\underbrace{0100000000}_{256}\underbrace{1011010001}_{721}\underbrace{1111111111}_{1023}\underbrace{1110000100}_{900} 
 \end{equation*}
 \item Element $M_{1,7}=700=1010111100$ does not fit on the remaining 4 bits of the first bitstring. So it will straddle two bitstrings, i.e., it is divided in two segments $a$ and $b$, $a$ is written on the first bitstring and $b$ on the second.
 \begin{equation*}\label{eq:07}
  \underbrace{\underbrace{0000010100}_{20}\overbrace{101011}^{b}}_{bitstring_2}|\underbrace{\overbrace{1100}^{a}\underbrace{0000001010}_{10}\underbrace{0000000001}_{1}\hdots}_{bitstring_1}
\end{equation*}
\end{enumerate}

 Please notice that bitstrings are written from right to left.

\begin{equation*}\label{eq:8}
  \underbrace{0000010100}_{20}\underbrace{\overbrace{101011}^a|\overbrace{1100}^{b}}_{700}\underbrace{0000001010}_{10}\ldots
\end{equation*}

Thus the compressed matrix $m = M_2$ requires less memory than the conventional storage of $M$ as a 64-bit integer array.

\subsubsection*{Method 2: Variable Length Blocks (VLB)}

In the SM method, there is still waste of space since for elements smaller than the supremum, a number of bits remain unused.

In the VBL method, the absolute minimal number of bits are used to store each value. However, if we are going to divide the biststrings into variable length chunks, we also need to reserve some extra bits to represent the size of each chunk, otherwise the elements cannot be recovered once they are stored.

Lets use again the matrix described in Equation \ref{eq:02}, where  the largest element is number 1023. Now instead of assigning one chunk of the bitstring to each element of $m$, we will assign two chunks: the first will store the number of bits required to store the element and the second  will store the actual element. The first chunk will have a fixed size, in this case, 4 bits. These 4 bits are the required space to store the bit-length of $\sup M $, in this case, 10. 

Lets go through VLB compression step-by-step. The largest element of $M$ is 1023. Its bit-length is 10 which in turn is 4 bits long in base 2 (1010). Thus the fixed size chunk is 4 bits long for every element.

\begin{enumerate}
 \item The first element $M_{1,1}=900$ requires 10 bits to store, so we write $10$ in the first chunk and $900$ in the second.
 \begin{equation*} \label{eq:09}
 bitstring_1 = 00000000000000000000000000000000000000000000000000\underbrace{\underbrace{1110000100}_{\text{element} = 900}\underbrace{1010}_{\text{bit-length=10}}}_{M_{1,1}}
\end{equation*}
 \item Do the same for the next element, $M_{1,2}=1023$.
  \begin{equation*} \label{eq:10}
 bitstring_1 = \ldots00000000000000000000000000\underbrace{\underbrace{1111111111}_{\text{element} = 1023}\underbrace{1010}_{\text{bit-length=10}}}_{M_{1,2}}\underbrace{\underbrace{1110000100}_{\text{element} = 900}\underbrace{1010}_{\text{bit-length=10}}}_{M_{1,1}}
\end{equation*}
 \item Element $M_{1,3}=721$ is also added taking the bitstring to the state.
 \begin{equation*} \label{eq:11}
 bitstring_1 = 
 0000000000000000000000\underbrace{1011010001}_{721}\underbrace{1010}_{10}\underbrace{1111111111}_{1023}\underbrace{1010}_{10}\underbrace{1110000100}_{900}\underbrace{1010}_{10}
\end{equation*}
\end{enumerate}

So far the VLB method is more wasteful than the SM, but when we add $M_{1,4} =256$  we start to save some space.

\begin{enumerate}
 \item[4.] Element $M_{1,4} =256$ is added.  
 \item [5.] Elements $M_{1,5} =1$ and $M_{1,6} =10$ are added requiring a total of 13 bits instead of 20 with the SM method. With the addition of these elements we require a second bitstring.
 \begin{align*} \label{eq:12}
 bitstring_1 &=
 \underbrace{0100}_{4}\underbrace{1}_{1}\underbrace{0001}_{1}\underbrace{100000000}_{256}\underbrace{1001}_{9}\underbrace{1011010001}_{721}\underbrace{1010}_{10}\underbrace{1111111111}_{1023}\underbrace{1010}_{10}\underbrace{1110000100}_{900}\underbrace{1010}_{10} \\
 bitstring_2 &= 000000000000000000000000000000000000000000000000000000000000\underbrace{1010}_{10}
\end{align*}
 \item [6.] The remaining two elements are added $M_{1,7} =700$ and $M_{1,8} =20$ in the second bit strip.
 \begin{equation*} \label{eq:13}
 bitstring_2 = 00000000000000000000000000000000000\underbrace{10100}_{20}\underbrace{0101}_{5}\underbrace{1010111100}_{700}\underbrace{1010}_{10}\underbrace{1010}_{10}
\end{equation*}
\end{enumerate}

 We used a total of 87 bits to store matrix $m$ with the VLB method instead of 80 bits using the SM method. However, as shall be seen later, the VLB method will be the most efficient for most matrices.
 
 \subsection*{Compression Efficiency}
 Compression efficiency depends of the data being compressed. Below, a formula for calculating compression efficiency is derived for both methods. They will be based on the following ratio:
 
 \begin{equation}\label{eq:14}
  \eta=\frac{bits\, alocated-bits\, used}{bits\,alocated}
 \end{equation}
 
 Where \textit{bits alocated} above mean total bits required for standard storage of the matrix, without compression, while bits used mean total bits requires to store the matrix after compression. From now on the efficiencies are denoted by $\eta_1$ for the SM method and by $\eta_2$ for the VLB method. 
 \subsubsection*{SM Method}
 Let $M_{r \times c}$ be the matrix we wish to compress. In comparison with a conventional allocation (64-bit integers), we can apply Equation \ref{eq:14} to caluculate the efficiency of the SM method:

\begin{align}\label{eq:15}
 \eta_1 &= \frac{64 \times rc - b(max M) \times rc}{64 \times rc}\nonumber\\
 &= \frac{64  - b(max M) }{64}
\end{align}

As we see in \ref{eq:15}, $\eta_1$ does not depend on size of the matrix, only on the bit-length of $\max M$. If $b(\max M)=64$, $\eta_1$ is 0, i.e., no compression is possible. On the other extreme, if the matrix is composed exclusively of 0s and 1s, maximal compression is achievable, $\eta_1=1$.
 
 \subsubsection*{VLB Method}
 
 For the VLB method, compression depends on the value of each element of the matrix. In this method bit-lentgh variability affects the compression ratio, so the formula will have to include this information.

Let the $rc$ elements of the matrix $M_{r \times c}$ be divided into $g$ groups, each with $f_i$ numbers of bit-length $b_i = b(m_i)$. Thus $f_i$ is the frequency of each bit-length present in $M$. Let $k = b(b(\max M))$, i.e., the bit length of the bit-length of $\max M$. The efficiency $\eta_2$ is shown below.

\begin{equation}\label{eq:16}
 \eta_2 = \frac{64 \times rc - \sum_{i=1}^{g} ( b_i + k ) \times f_i }{64 \times rc} 
\end{equation}

We can further simplify Equation \ref{eq:16} to get at shorter expression for the compression ratio.

\begin{align}
 \eta_2 &= \frac{64 \times rc - \sum_{i=1}^{g} ( b_i \times f_i + k \times f_i )}{64 \times rc} \nonumber \\
  &= \frac{64 \times rc - \sum_{i=1}^{g}  b_i \times f_i  -\sum_{i=1}^{g}  k \times f_i }{64 \times rc}\nonumber \\
  &= \frac{64 \times rc - \sum_{i=1}^{g}  b_i \times f_i  - k \times\sum_{i=1}^{g}  f_i }{64 \times rc}\nonumber 
\end{align}
  
 Knowing that $$\sum_{i=1}^{g} f_i = rc, $$ we can simplify the equation above, obtaining (\ref{eq:17}).
 
\begin{align}\label{eq:17}
 \eta_2 &= \frac{64 \times rc - \sum_{i=1}^{g}  b_i \times f_i  - k \times rc }{64 \times rc}\nonumber \\
  &= 1 - \frac{\sum_{i=1}^{g}  b_i \times f_i }{64 \times rc} - \frac{k}{64}
\end{align}

% Results and Discussion can be combined.
\section*{Results}

\paragraph{Random Matrix Generation.}
In both methods, compression efficiency depends on the distribution of the bit-lengths $b(m_{i,j})$. Thus, in this section, a method to generate a variety of random bit-length distributions is proposed.

For simplicity we will model the distribution of  as a mixture $X$ of two Beta distributions, $B_1 \sim Beta(\alpha_1,\beta_1)$ and $B_2 \sim Beta(\alpha_2,\beta_2)$, whose probability function is shown in Equation \ref{eq:18}. Since the Beta distribution is defined only in the interval $[0,1] \subset \mathbb{R}$ , we applied a simple transformation ($\lfloor 64 \times x \rfloor + 1$) to the mixture in order to map it to the interval of $[1,64] \subset \mathbb{Z}$.

\begin{equation}\label{eq:18}
 F(x) =  w \, Beta(\alpha_1,\beta_1) + (1-w)\,Beta(\alpha_2,\beta_2)
\end{equation}

The intention of using this mixture was to find a simple way to represent a large variety of bit length distributions. The first two central moments of this mixture are given in \ref{eq:19} and will be used later to summarize our numerical results.

\begin{align}\label{eq:19}
 E(X) &= w E(B_1) + (1-w) E(B_2) \nonumber \\
 &= w \frac{\alpha_1}{\alpha_1+\beta_1} + (1-w)\frac{\alpha_2}{\alpha_2+\beta_2}\nonumber \\
 Var(X) &= w Var(B_1) + (1-w) Var(B_2) + w (1-w) (E(B_1)^2 - E(B_2)^2)
\end{align}

In order to explore the compression efficiency of both methods, we generated samples from the mixture defined above, varying its parameters. From now on, when we mention Beta distribution we will mean the transformed version defined above.

From now on we will apply Equations \ref{eq:15} and \ref{eq:17}, to determine the compression efficiency of SM and VLB methods for random matrices generated as describe above. 

\begin{figure}[ht]
  \centering
  \subfigure[$\alpha=1,\beta_1=1$]{
  \includegraphics[scale=0.25,angle=-90,clip]{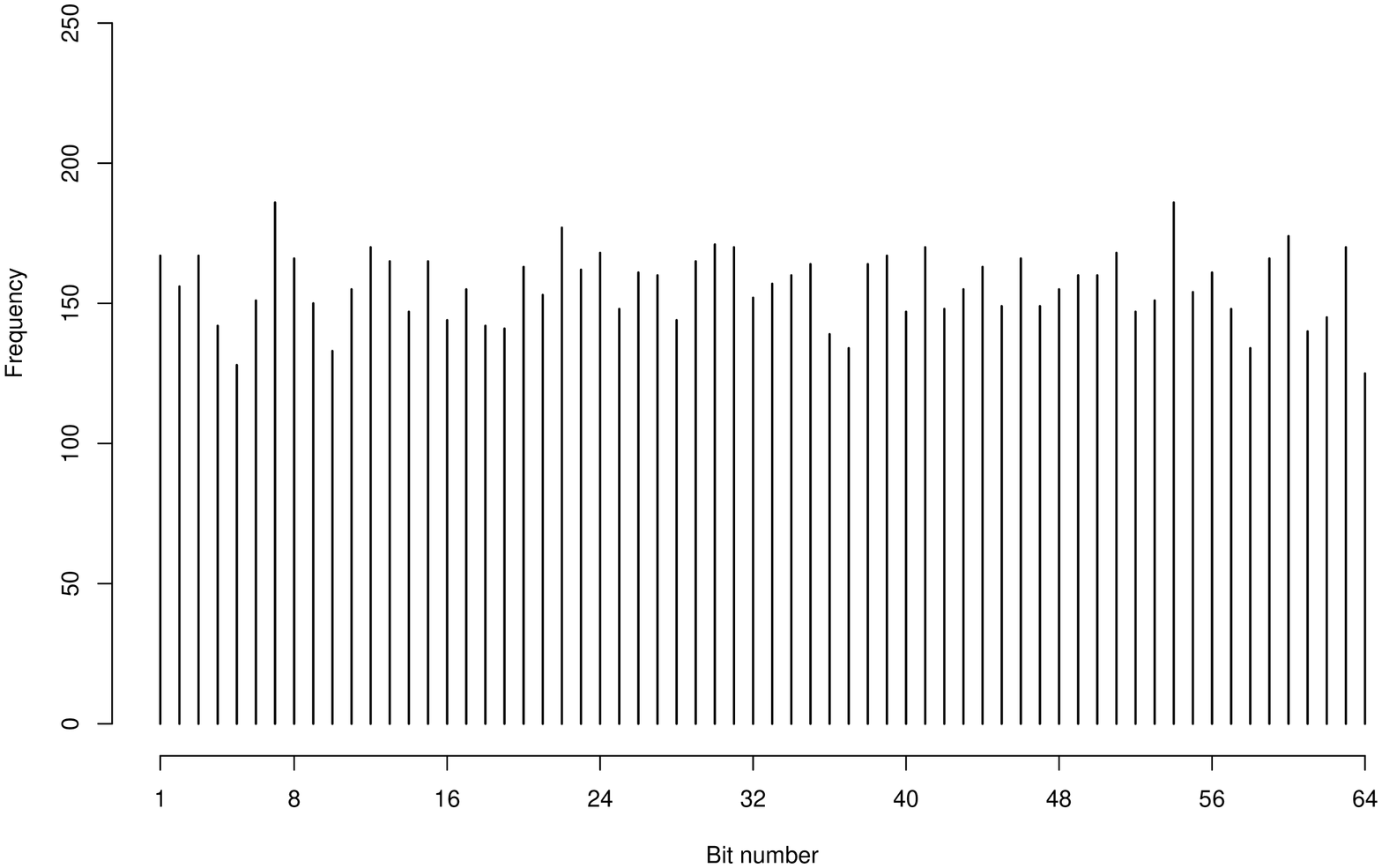}
  }
  \subfigure[$\alpha=1,\beta=32$]{
  \includegraphics[scale=0.25,angle=-90,clip]{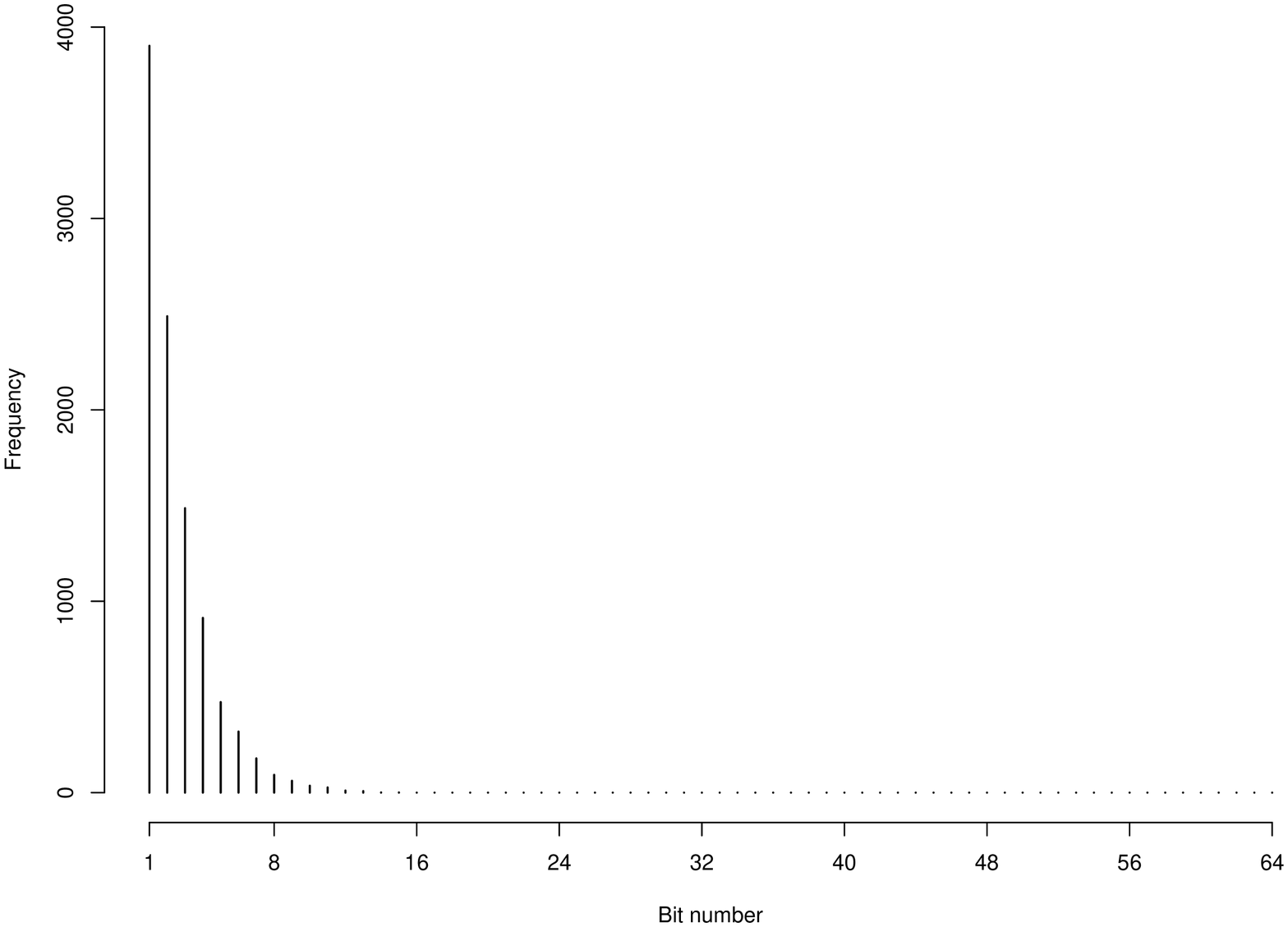}
  }
  \subfigure[$\alpha=32,\beta=1$]{
  \includegraphics[scale=0.25,angle=-90,clip]{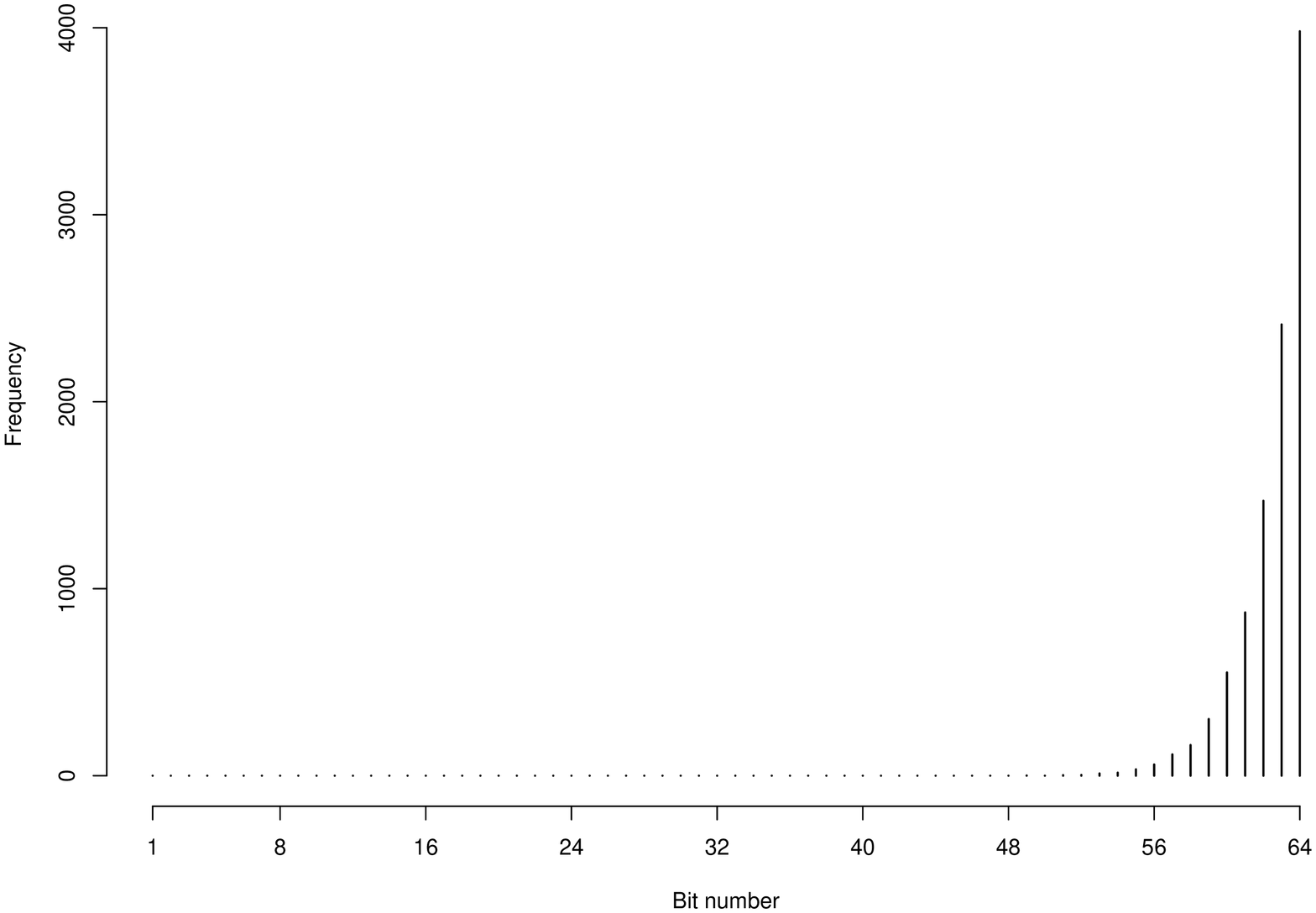}
  }
  \subfigure[$\alpha=64,\beta=64$]{
  \includegraphics[scale=0.25,angle=-90]{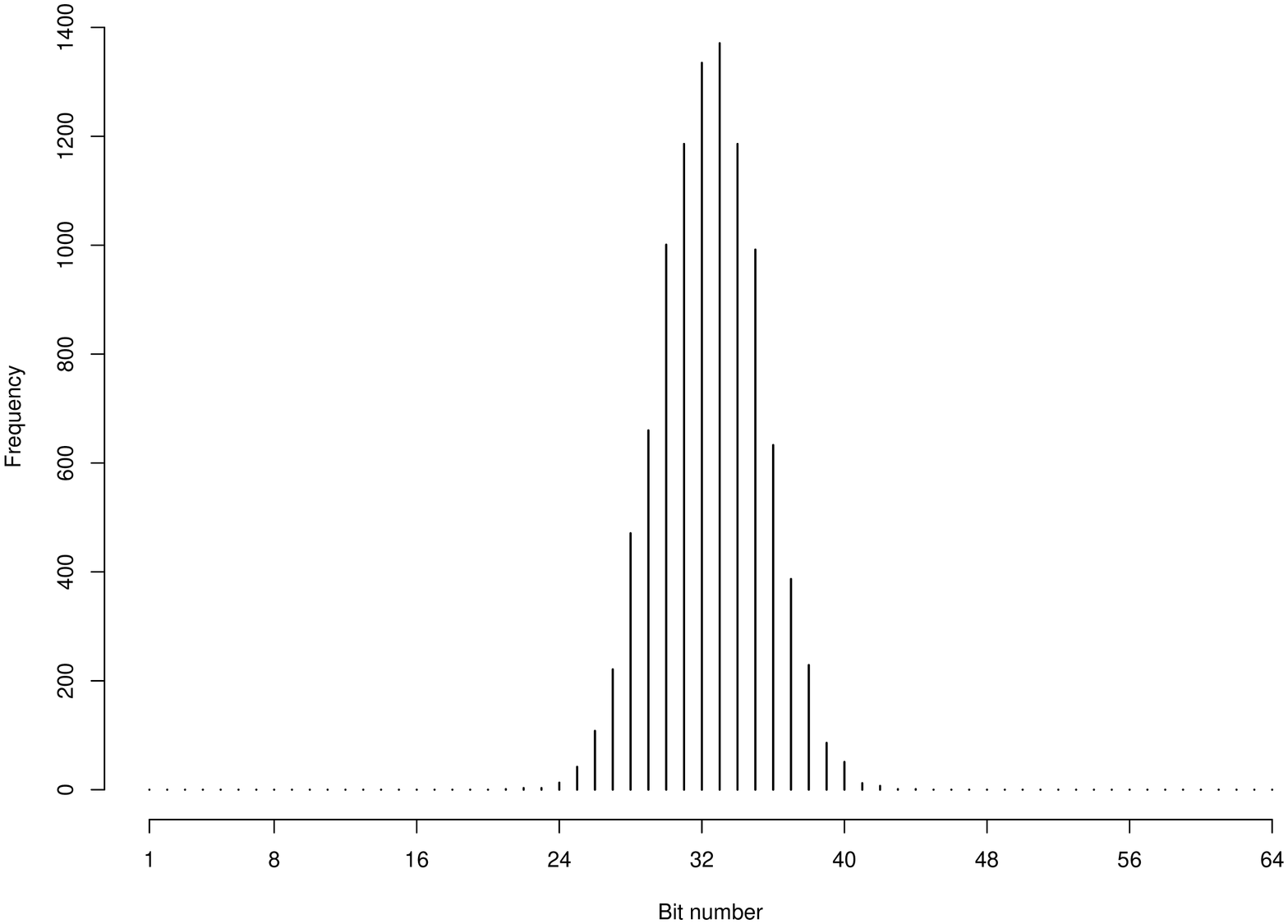}
  }
  \caption{Histograms constructed from samples with 10,000 elements, generated from a Beta distribution. Below each histogram is possible to verify the parameters used.}
  \label{fig:01020304}
\end{figure}

With $w=0$, a single Beta distribution is used. In Figure \ref{fig:01020304}, we show some distributions of bit-lengths for some combinations of $\alpha_1$ and $\beta_1$. From the figure it can be seen that a large variety of unimodal distributions can be generated in the interval $[1,64]$.

As we are sampling from a large set distributions of bit-length, represented by the mixture of betas presented above, in order to make our results more general, we will base our analysis on the expected bit-length of a sample, since the efficiency of both methods depends on it. So, from Equations \ref{eq:15} and \ref{eq:17}, the expected efficiencies become: 
 
\begin{equation}\label{eq:20}
 E(\eta_1) = 1 - \frac{k}{64}
\end{equation}

\begin{equation}\label{eq:21}
 E(\eta_2) = 1 - \frac{E(b)}{64} - \frac{k}{64}
\end{equation}

\noindent where k, in (\ref{eq:21}), is set to 7 (the bit-length required to represent the largest possible bit-length: 64). In (\ref{eq:20}), $k$ is the bit-length of the greatest element, or in the worst case, 64.

We will use the difference $D=E(\eta_1)-E(\eta_2)$ to compare the efficiency of the two methods. Thus a positive $D$ will favor SM method while a negative $D$ favors VLB method.

The expected compression efficiency in the following numeric experiments, will be calculated from $3$ matrices of dimension $10000$, generated as described, and presented in tables and figures below. 
 
In Figure \ref{fig:06070809}, we can see the distribution of efficiencies and their difference for a sample generated from a single Beta distribution of bit-lengths. Note that both methods can achieve efficiencies greater than 80\% for matrices with very small numbers. Also note that the VLB method is more efficient in the majority of cases.
 
\begin{figure}[h]
  \centering
   \subfigure[SM-VLB  Difference ($D = \eta_1 - \eta_2$)]{ 
   \includegraphics[scale=0.18,clip]{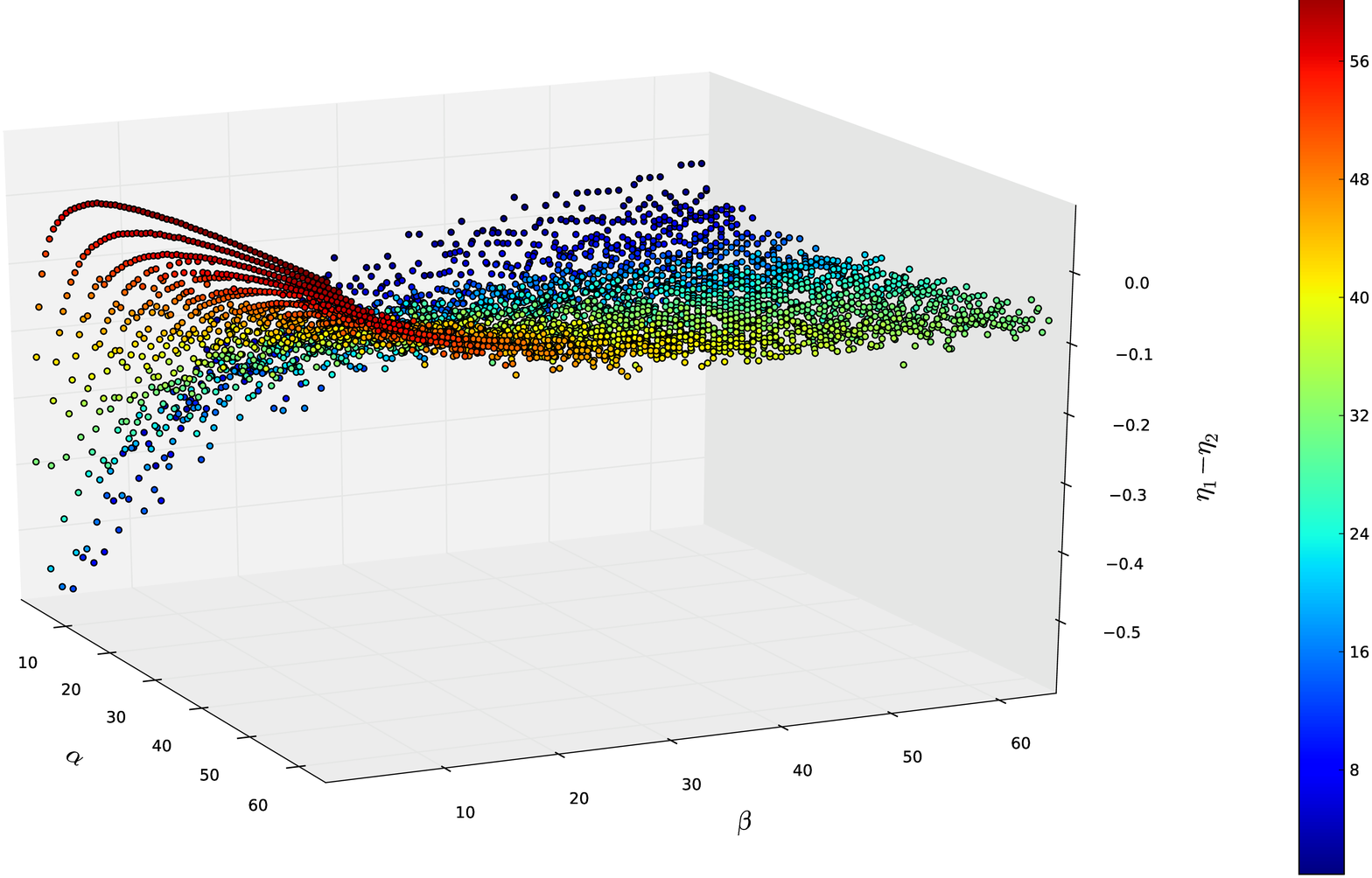}
   }
   \subfigure[SM Efficiency  ($\eta_1$)]{
   \includegraphics[scale=0.18,clip]{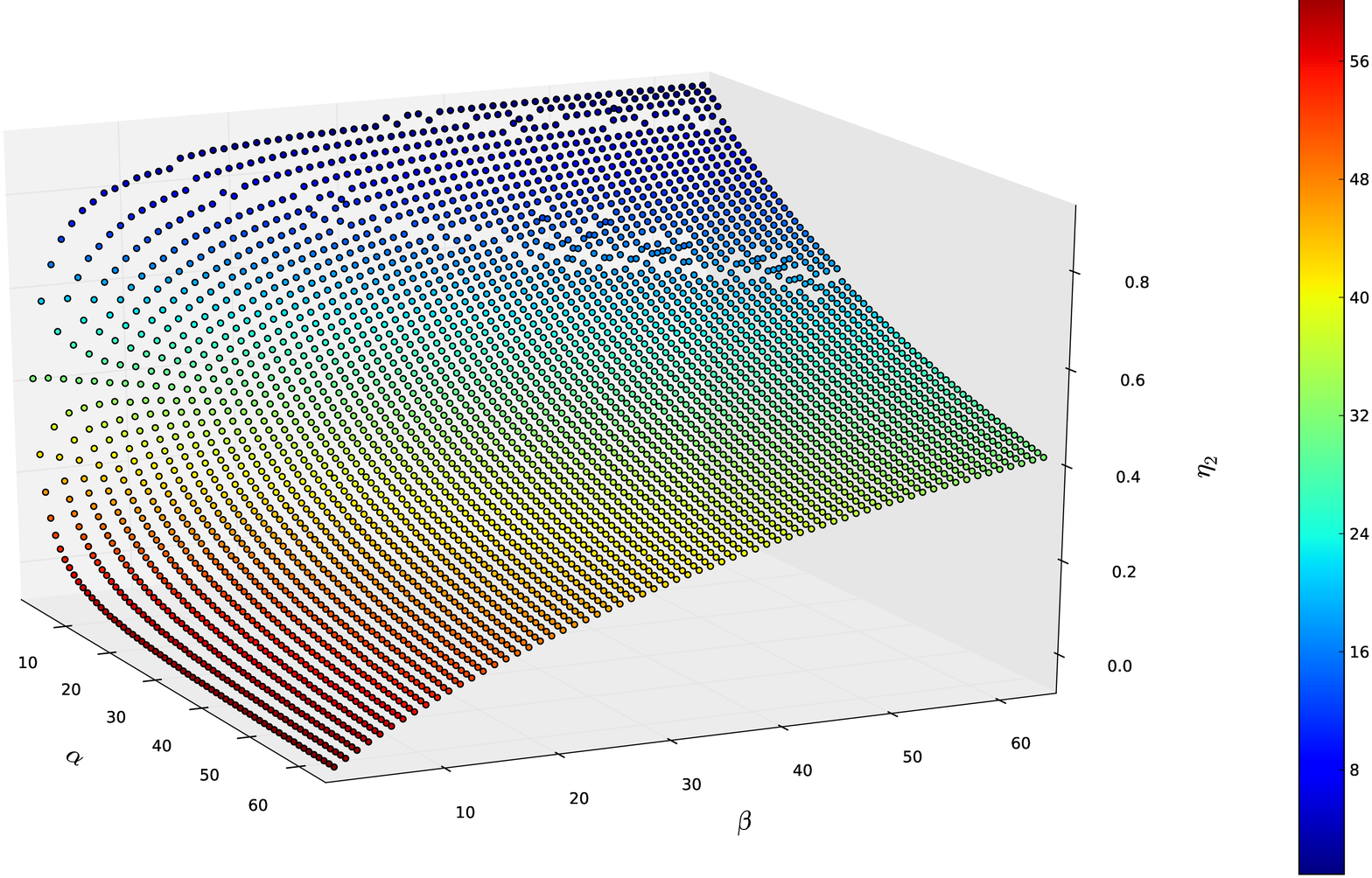}
   }
   \subfigure[VBL Efficiency ($\eta_2$)]{
   \includegraphics[scale=0.18,clip]{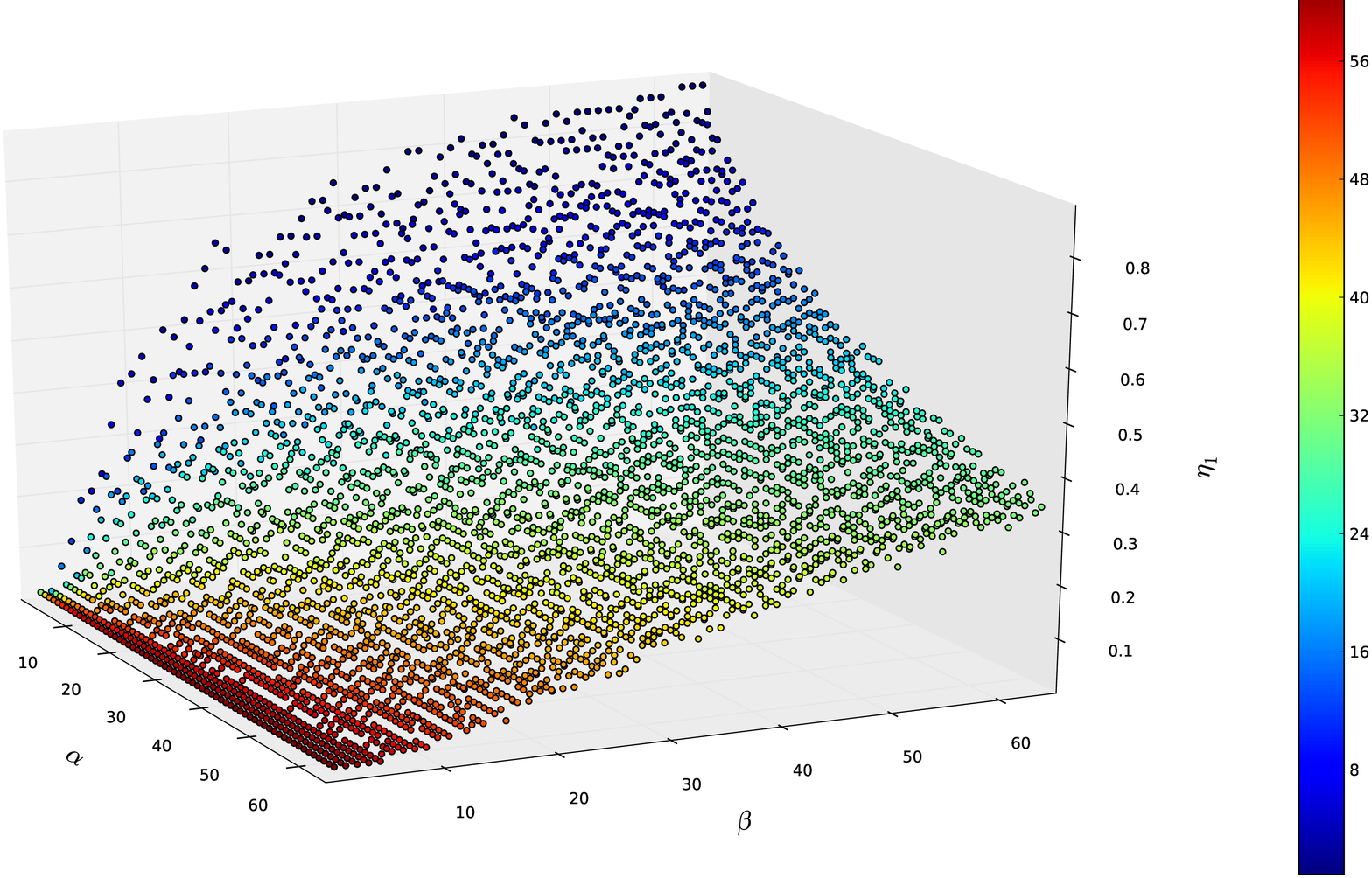}
   }
   \subfigure[SM$\geq$VLB Efficiency ($\eta_1 \geq \eta_2$)]{
   \includegraphics[scale=0.18,clip]{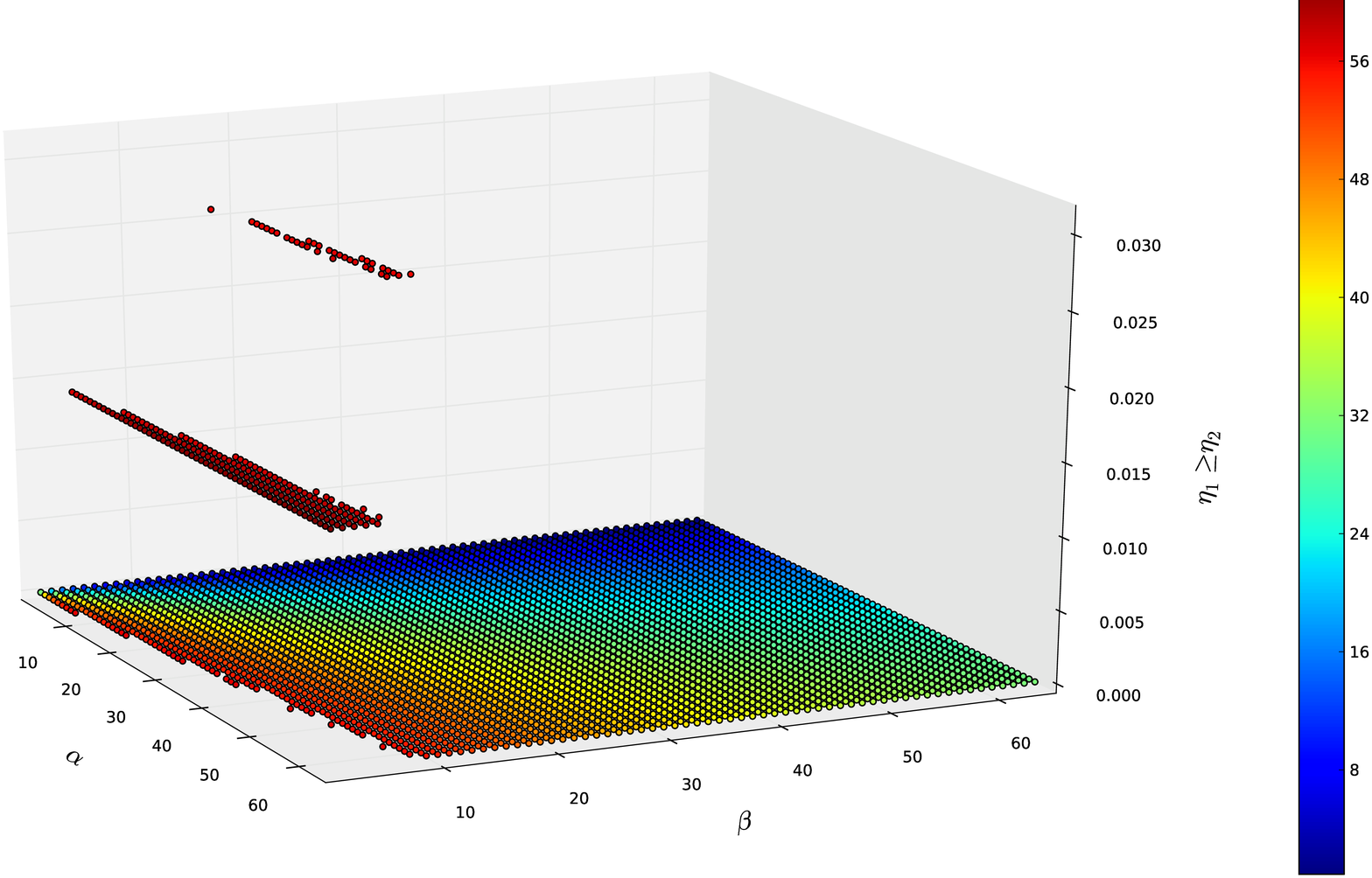}
   }
  \caption{Comparing compression efficiency of methods 1 and 2. Color scale in (a), (b) and (c) represent average bit-length. In (a) we can see the difference $D = \eta_1 - \eta_2$. It can be seen that for most combination of $\alpha$ and $\beta$, $D<0$, meaning the second method is more efficent to compress a sample of numbers with bit-lengths coming from a $Beta(\alpha,\beta)$ distribution. However, there is a small region in parameter space, which is shown in white on (d), where the SM method is more efficient. This region corresponds to the dots in red in (a), where the average bit-length is higher. In panels (b) and (c), we can see the efficiencies of SM and VLB methods, respectively.}
  \label{fig:06070809}
\end{figure}
 
Now let $w = 0.5$, i.e., matrices will have elements with bit-lengths comming from a mixture of beta distributions, $B_1\sim Beta(\alpha_1,\beta_1)$ and $B_2\sim Beta(\alpha_2,\beta_2)$. The expected value for this mixture is shown in Equation \ref{eq:22}.
 
\begin{equation}\label{eq:22}
 E(B) = 0.5 E(B_1) + 0.5 E(B_2)
\end{equation}
 
For bit-lengths coming from a mixture ($w>0$), let the expected efficiencies for the SM and VLB methods be as given by Equations \ref{eq:23} and \ref{eq:24}. So now, instead of having the efficiency be a function of greatest bit-length in the sample (denoted as  $k$ in \ref{eq:15} and \ref{eq:17}), it will be a function of $max\{E(B_1),E(B_2)\}$.

\begin{equation}\label{eq:23}
 E(\eta_1) = 1 - \frac{max\{E(B_1),E(B_2)\}}{64}
\end{equation}

\begin{equation}\label{eq:24}
 E(\eta_2) = 1 - 0.5\frac{E(B_1)}{64} - 0.5\frac{E(B_2)}{64} - \frac{max\{E(B_1),E(B_2)\}}{64}
\end{equation}
 
As before, we generate 3 matrices of dimension $10,000$ for each parameterization, calculate the average efficiencies (Equations \ref{eq:23} and \ref{eq:24}) and their diference $D$.

\begin{figure}[h]
  \centering
  \subfigure[$\alpha_1=1,\beta_1=1,\alpha_2=1,\beta_2=1$]{
  \includegraphics[scale=0.28,angle=-90,clip]{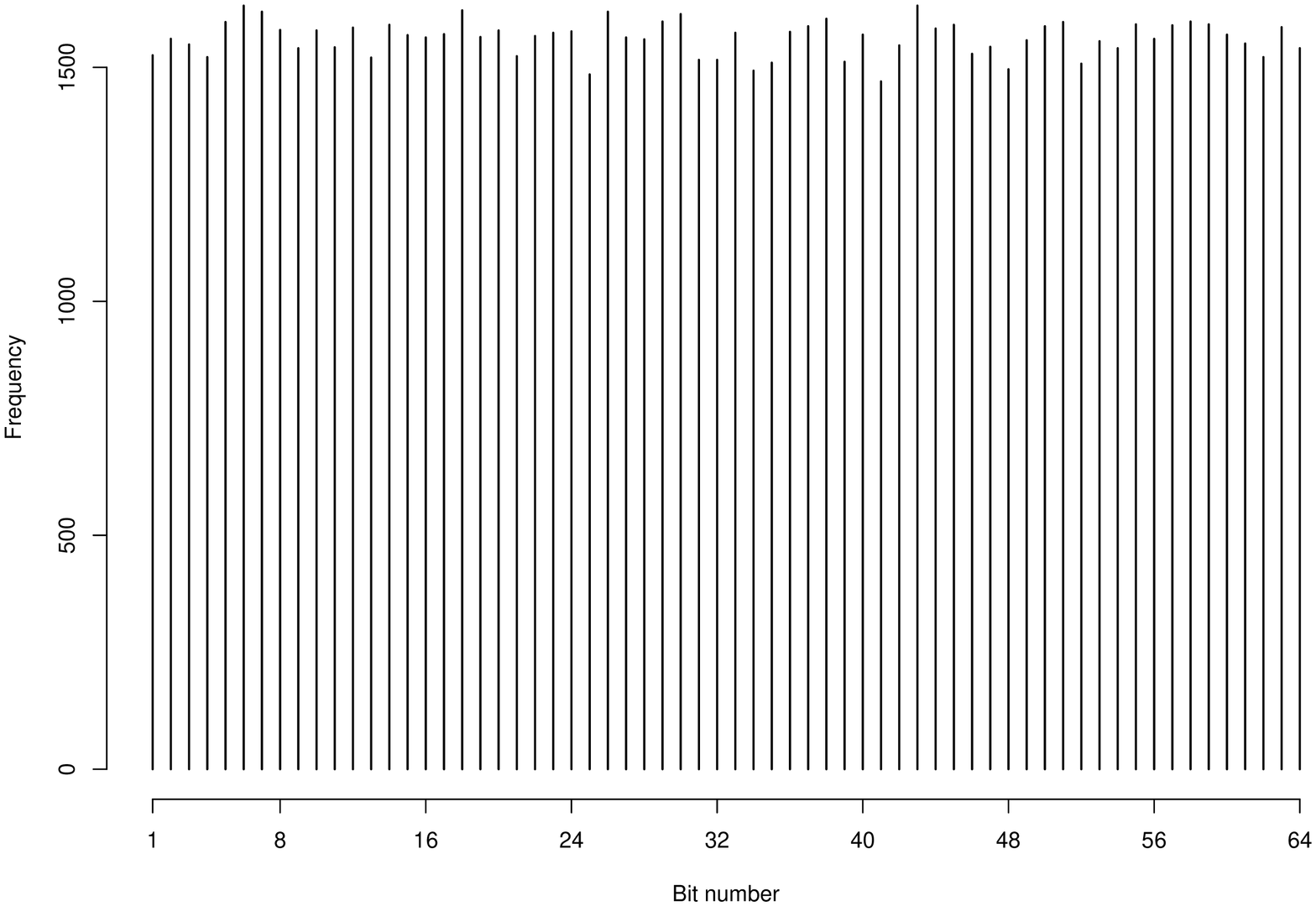}
  }
  \subfigure[$\alpha_1=1,\beta_1=32,\alpha_2=32,\beta_2=1$]{
  \includegraphics[scale=0.28,angle=-90,clip]{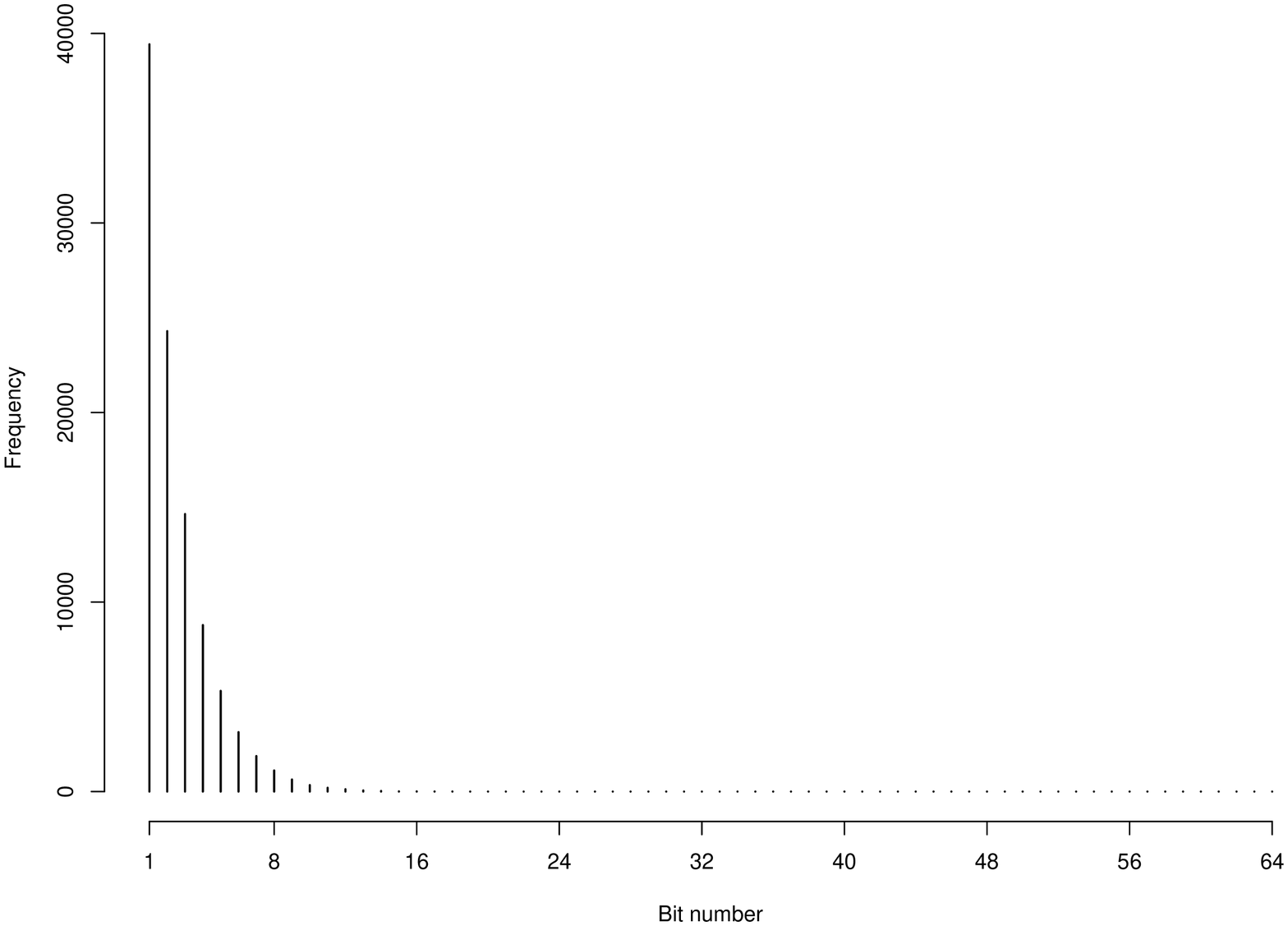}
  }
  \subfigure[$\alpha_1=32,\beta_1=32,\alpha_2=32,\beta_2=32$]{
  \includegraphics[scale=0.28,angle=-90,clip]{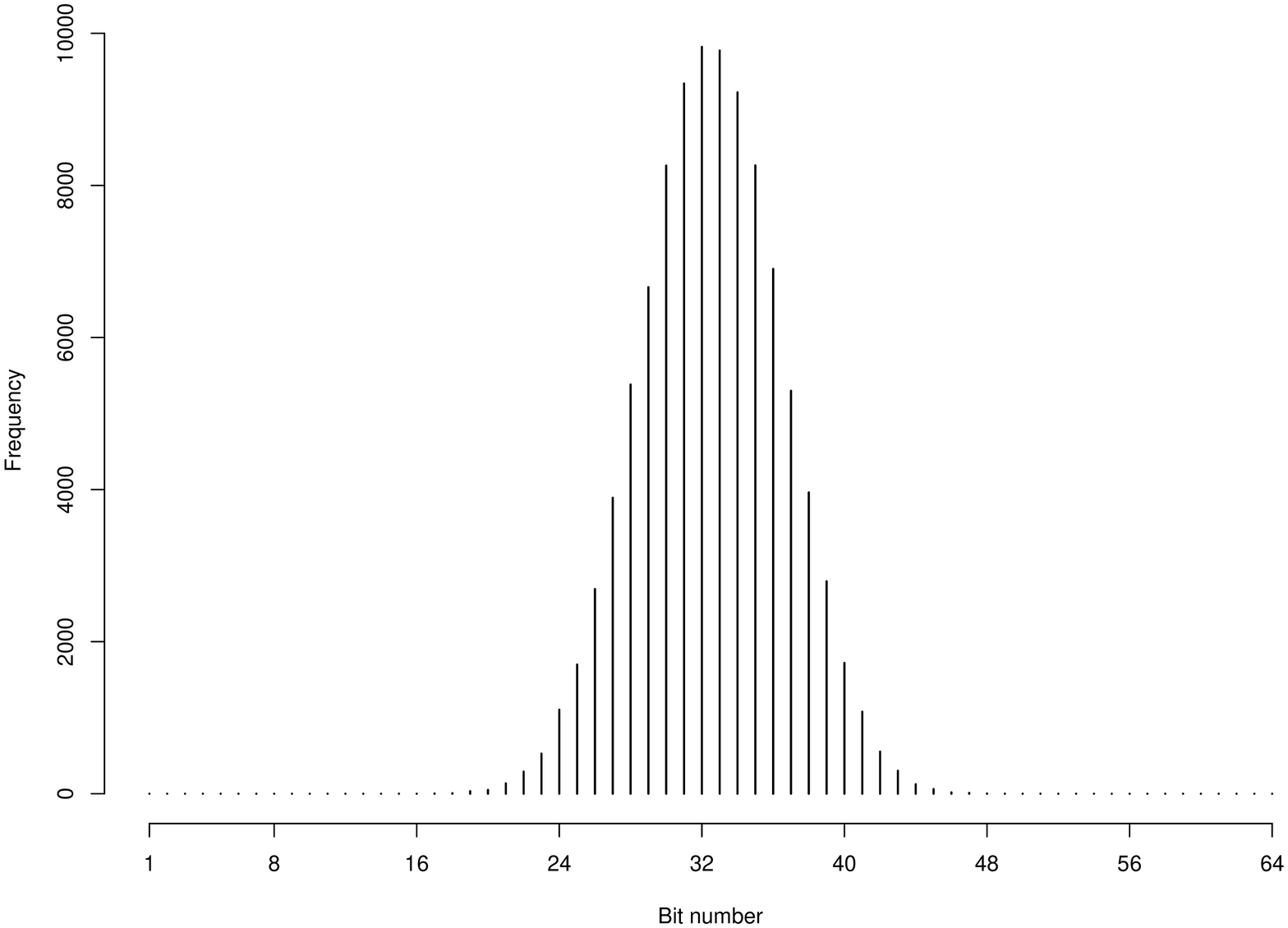}
  }
  \subfigure[$\alpha_1=64,\beta_1=32,\alpha_2=32,\beta_2=64$]{
  \includegraphics[scale=0.28,angle=-90]{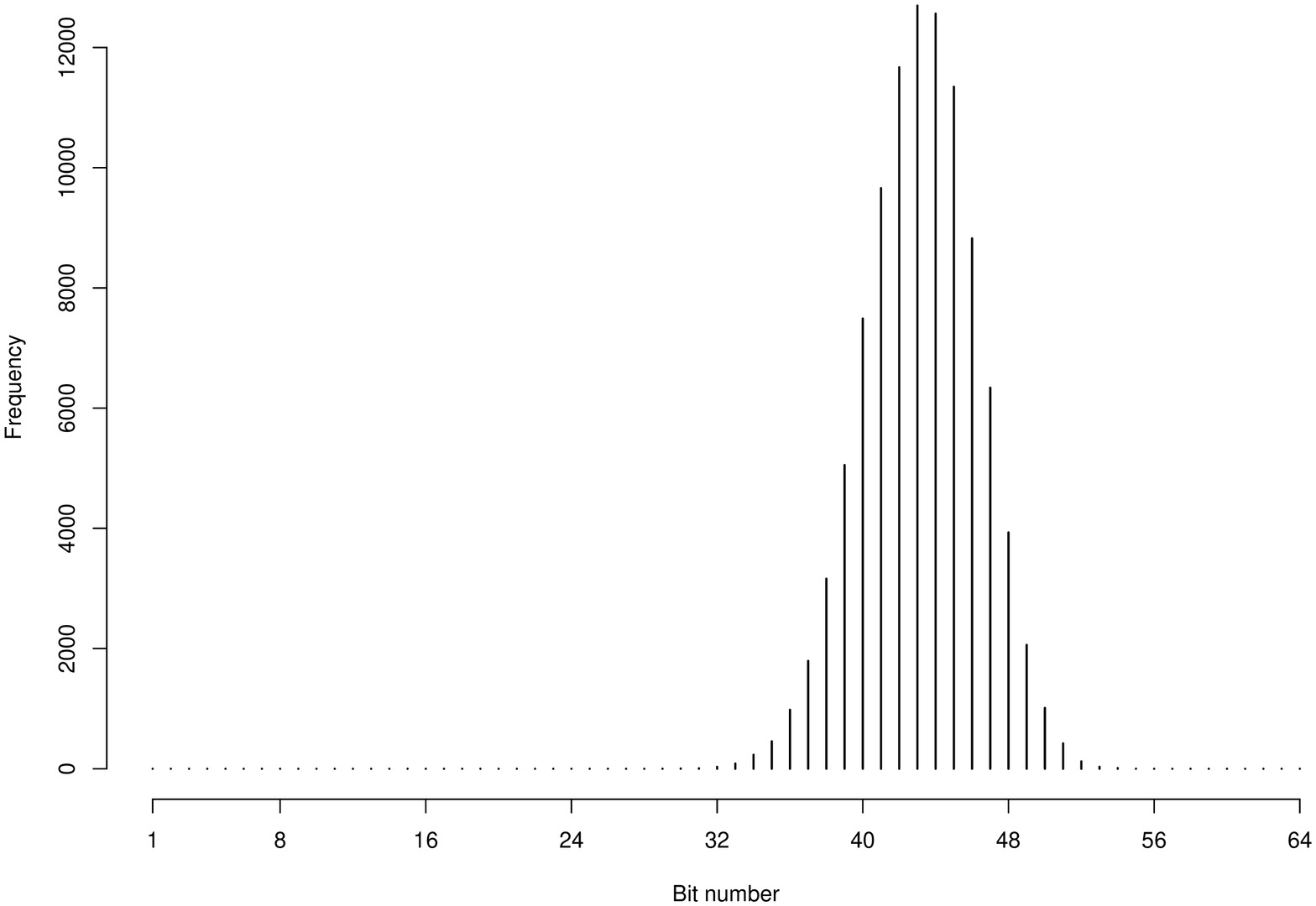}
  }
  \caption{Histograms constructed from samples with 10,000 elements, generated from the mixture of two Beta distributions with $w=0.5$. Below each histogram are the parameters of the mixture.}
  \label{fig:09101112}
\end{figure}

\begin{figure}[h]
  \centering
  \subfigure[$\alpha_1=64,\beta_1=48,\alpha_2=1,\beta_2=48$]{
  \includegraphics[scale=0.28,angle=-90]{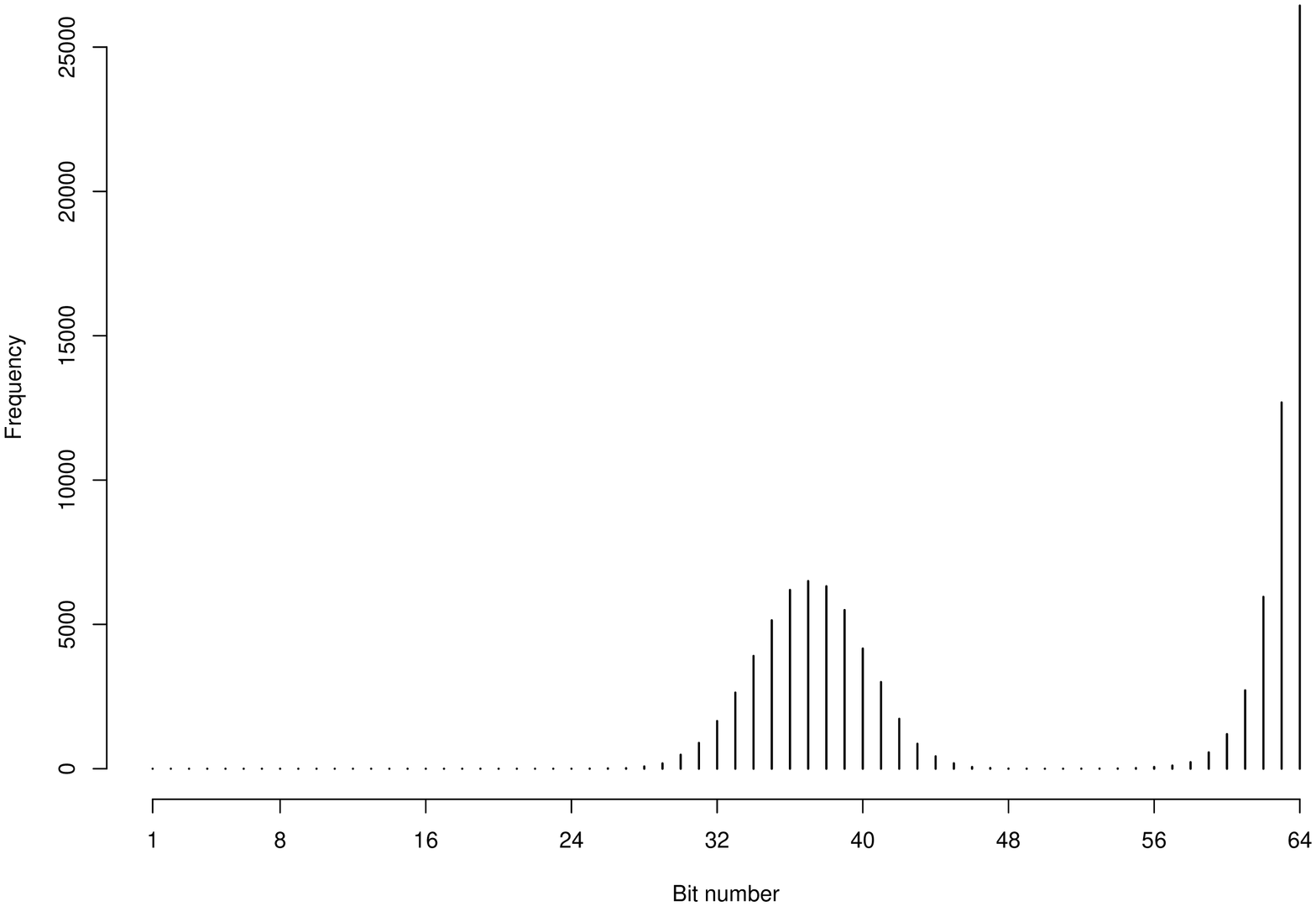}
  }
  \subfigure[$\alpha_1=16,\beta_1=46,\alpha_2=49,\beta_2=64$]{
  \includegraphics[scale=0.28,angle=-90]{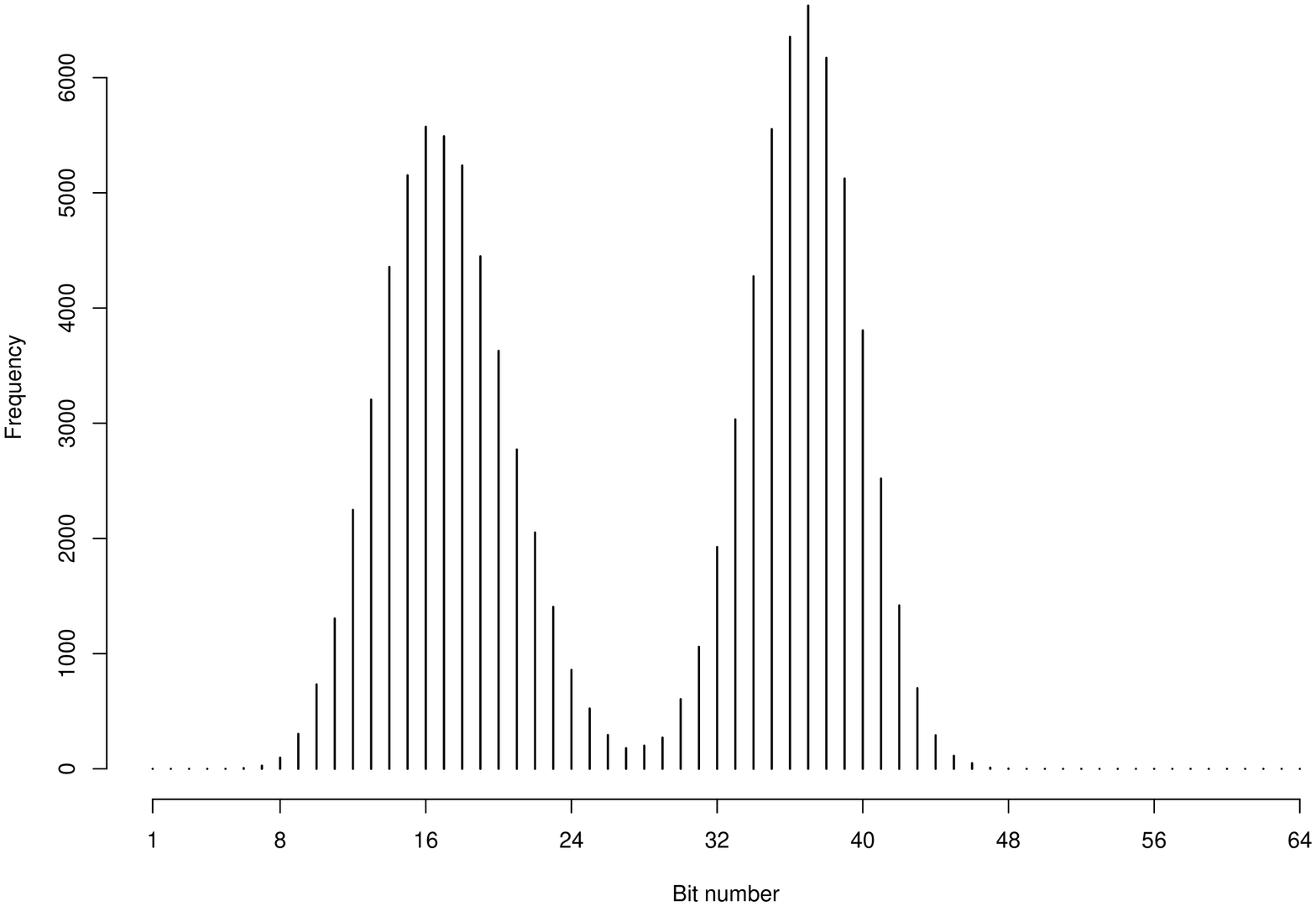}
  }
  \subfigure[$\alpha_1=16,\beta_1=16,\alpha_2=16,\beta_2=49$]{
  \includegraphics[scale=0.28,angle=-90]{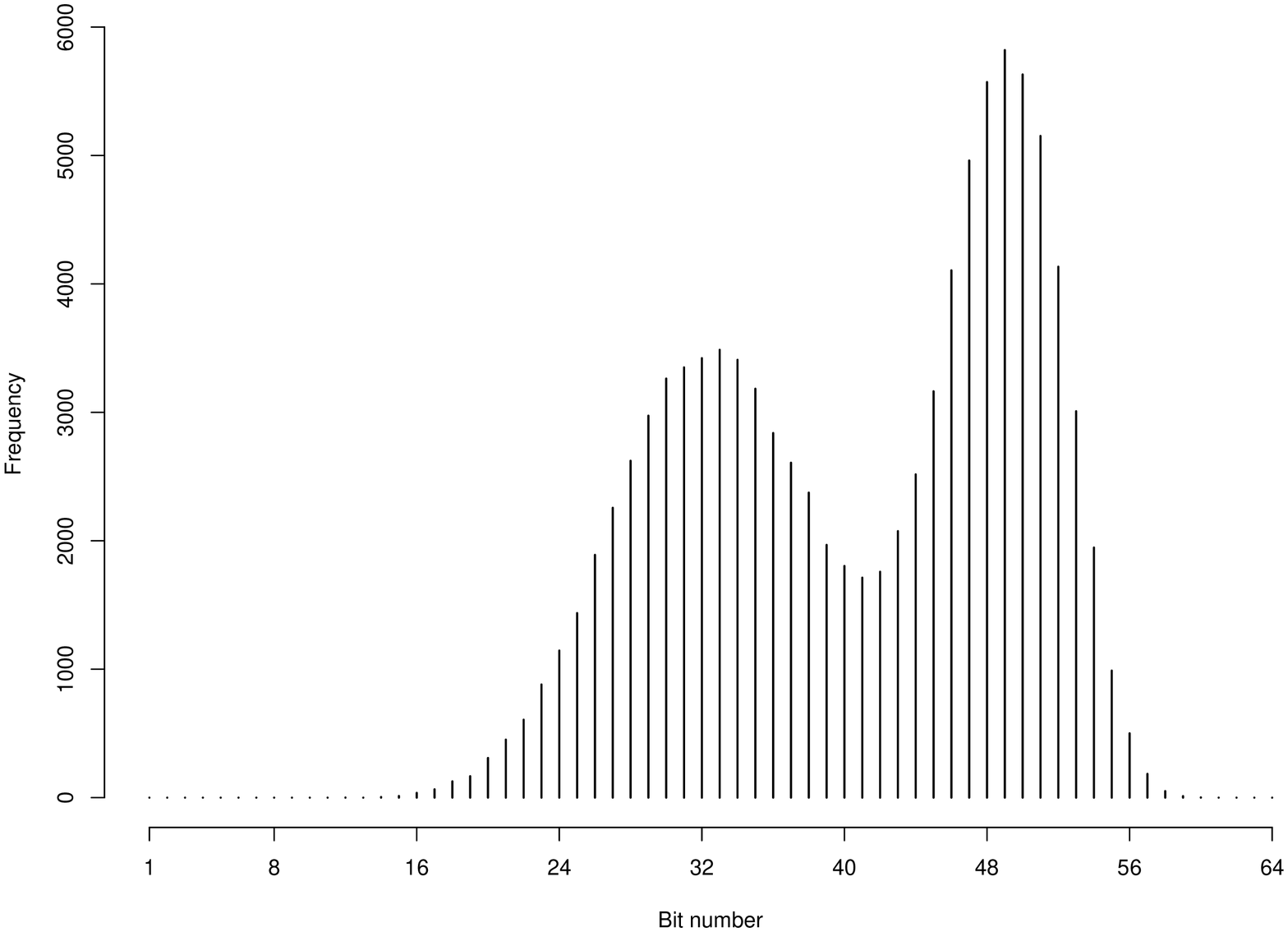}
  }
  \subfigure[$\alpha_1=1,\beta_1=16,\alpha_2=1,\beta_2=49$]{
  \includegraphics[scale=0.28,angle=-90]{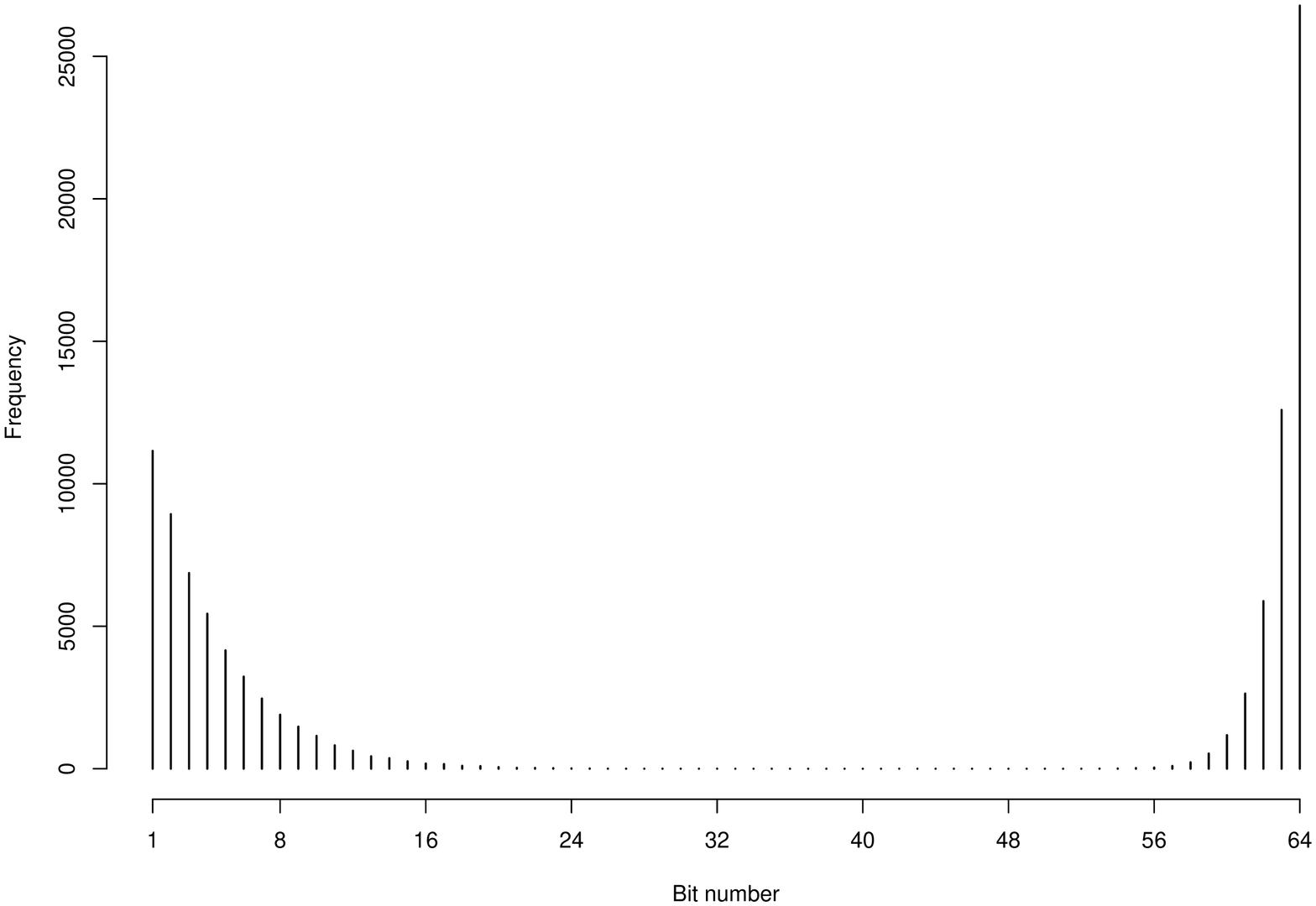}
  }
  \caption{Histograms constructed from samples with 10,000 elements, generated from the mixture of two Beta distributions with $w=0.5$. Below each histogram are the parameters of the mixture.}
  \label{fig:13141516}
\end{figure}

Before moving on to efficiency results and analyses, let's first inspect samples from the mixture of transformed Beta distributions. Figures \ref{fig:09101112} and \ref{fig:13141516}, show a few parameterizations and their resulting sample distributions. It is important to note that from the mixture we can now generate bimodal distributions as well as the unimodal types tested before. Since we are making statements about efficiency as a function of the expected bit-length, it is important to verify if these statements hold for bimodal  distributions as well.

\begin{figure}[h]
  \centering
  \subfigure[Efficiency histogram of the SM method ]{
  \includegraphics[scale=0.25,angle=-90]{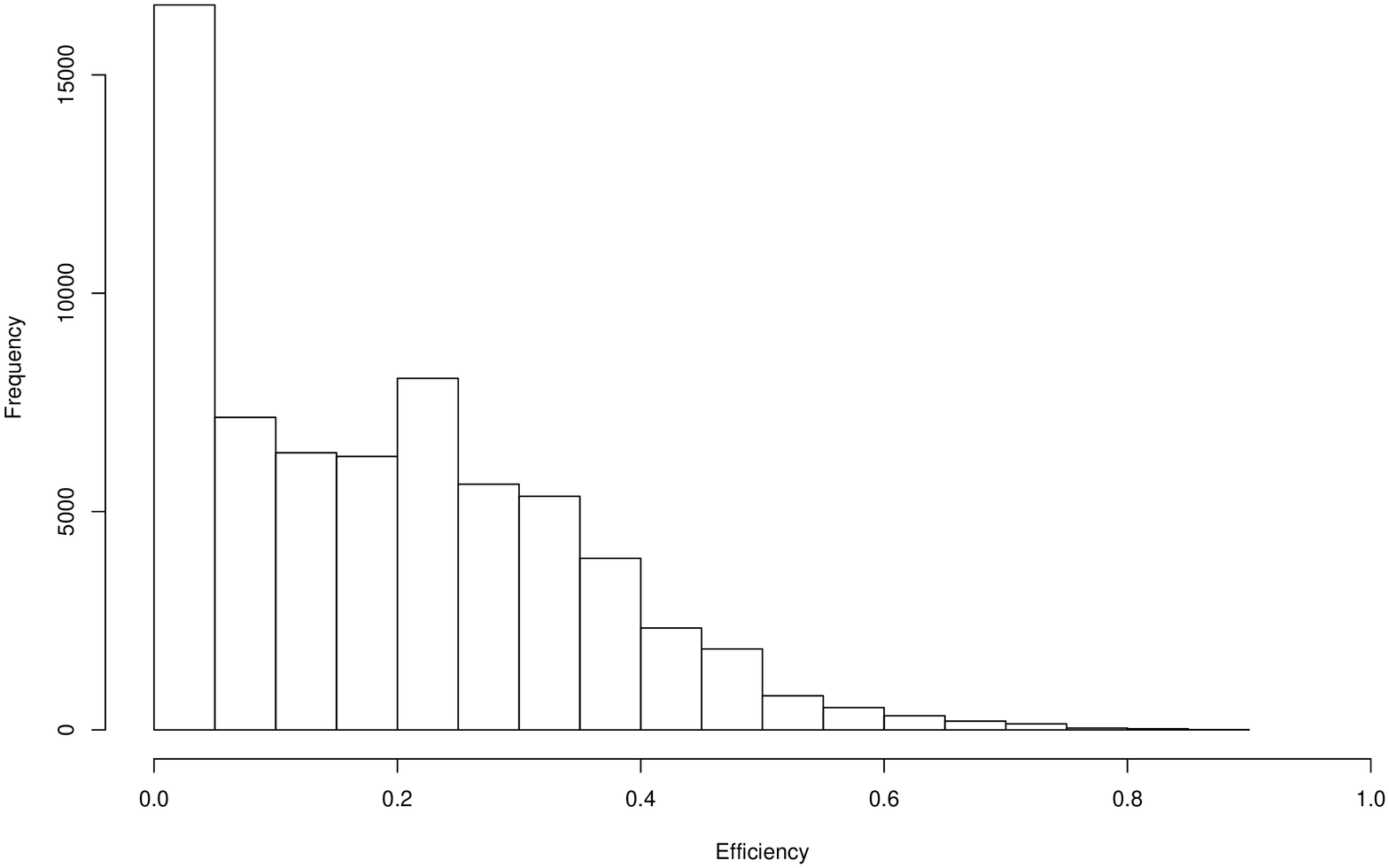}
  \label{fig:17}
  }
  \subfigure[Efficiency histogram of the VLB method ]{
  \includegraphics[scale=0.25,angle=-90]{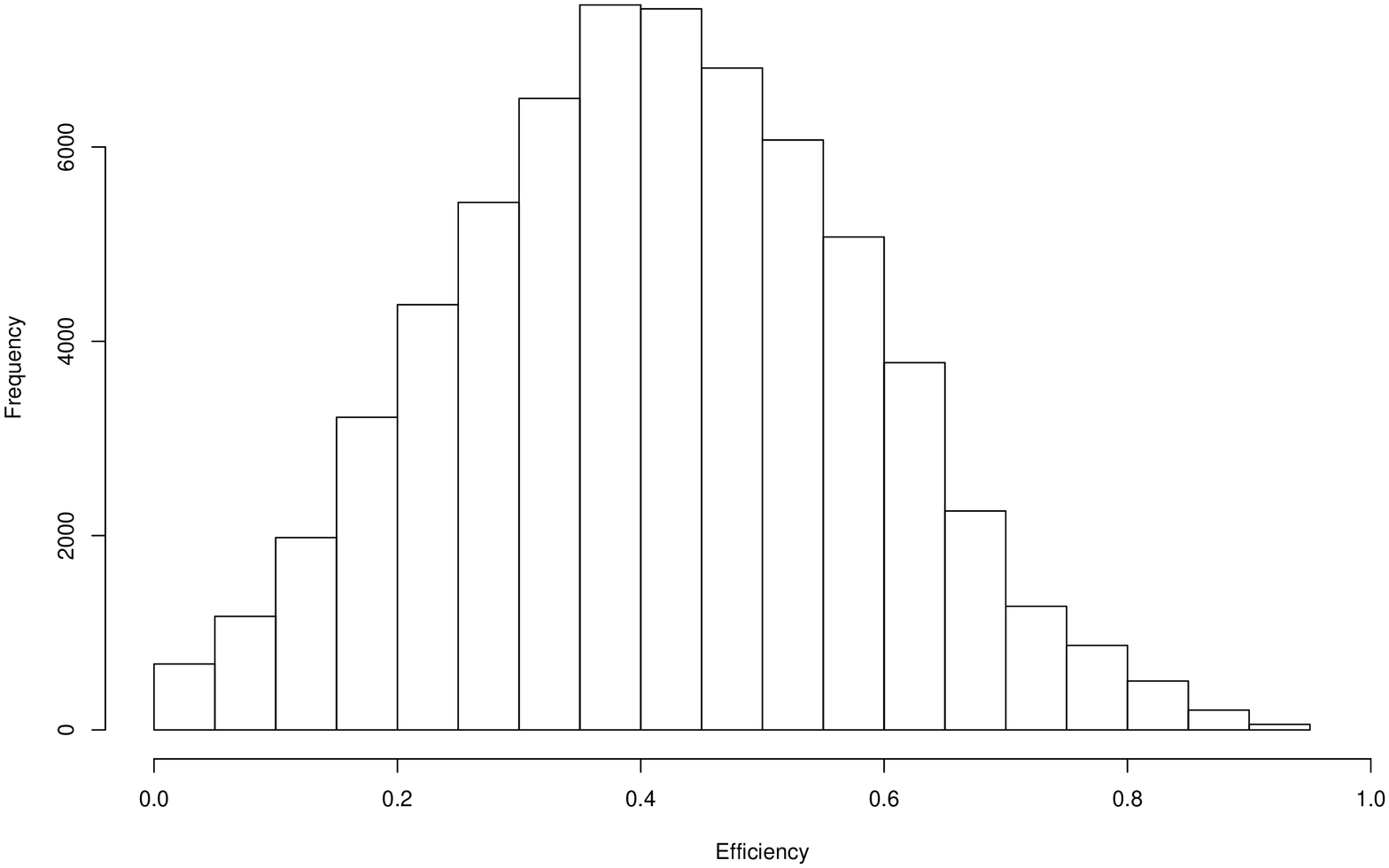}
  \label{fig:18}
  }
  \caption{Efficiency histograms of the SM and VLB methods. Note that the VLB method has a greater average efficiency than SM method .}
  \label{fig:1718}
\end{figure}

After sampling uniformly ($[1,5,9,\ldots,64]$, $n=65,536$) the bit-length space and comparing efficiencies, we summarized the results on Table \ref{tab:02}. In it we see how many parameterizations (from our sample) favor each method. We can also look at the distribution of efficiencies on our samples for each method (Figure \ref{fig:1718}), which clearly demonstrate the greater expected efficiency of method VLB (Figure \ref{fig:18}).

\begin{table}[h]
 \centering
 \caption{Efficiency comparison of SM and VLB methods for parameters covering uniformly the support of $B$. Column $n$ shows the number of parameter combinations with which each method has superior compression.}
 \begin{tabular}{ccc}
  \hline 
  Methods  & n   & Percentage \\
  \hline
  SM	   & 592	& 0.9034\% \\
  VLB	   & 64944	& 99.0966\% \\
  \hline
  Total    & 65536	& 100\% \\
  \hline
 \end{tabular}
 \label{tab:02}
\end{table}

\begin{figure}[h]
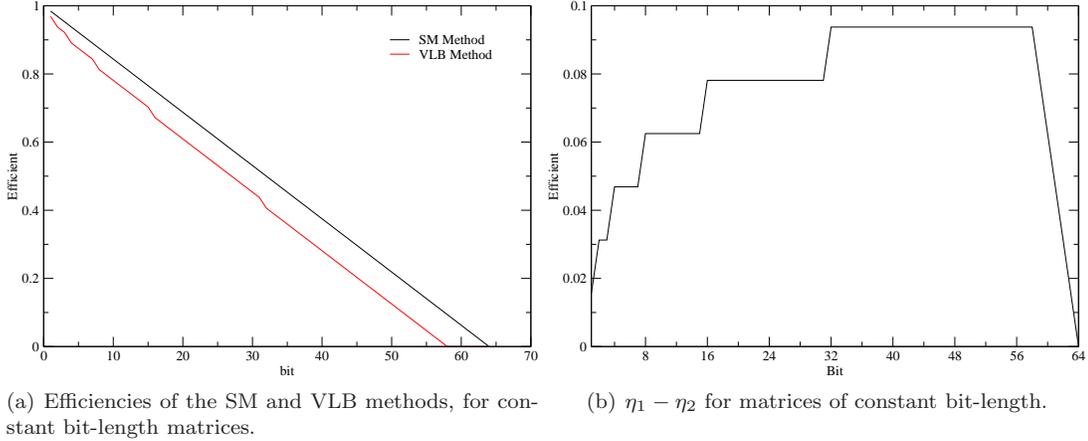

  \centering
  \subfigure[Efficiencies of the SM and VLB methods, for constant bit-length matrices.]{\includegraphics[scale=0.3,clip]{fig19}\label{fig19}}
  \subfigure[$\eta_1-\eta_2$ for matrices of constant bit-length.]{\includegraphics[scale=0.3,clip]{fig20}\label{fig20}}
  \caption{Compression efficiency of the SM and VLB methods for matrices of constant bit-length.}
  \label{fig:1920}
\end{figure}

As we have shown, the VLB method is more effective compressing most integer datasets up to 64 bits in size. This is due to its ability to exploit the variance in the data set and reduce the waste of bits in the representation of some numbers. In specific cases where the variance in the data null or too small, method I will be more efficient. As a matter of fact, for matrices where all elements have the same bit-length, SM method  will always be better, regardless of bit-length (Figures \ref{fig19} and \ref{fig20}), The only exception if for bit-length 64 where neither method is able to compress the data. 

\section*{Discussion}

\subsection*{Calculating Efficiencies}

To determine the best compression method to apply, it's necessary to inspect the distribution of bit-lengths of matrix elements. When matrix elements are small or have nearly-constant bit-length, the SM Method is best, otherwise, the VLB method should be chosen.

As an example, let $M_{r\times c}$ be a integer matrix such that the half of its elements have bit-length $1$ and the other half $64$. Recalling Equation \ref{eq:17}, now we have two groups of elements (by bit-length), $b_1=1 $, $b_2=64$ and $f_i = \frac{rc}{2}$ for $i = 1 \text{ and } 2$. As the greatest bit-length is $64$, then $k=7$. Compression efficiency $\eta_2$ can be calculated using Equation \ref{eq:17}. After plugging in our numbers, we obtain a compression of $38.29\%$.

\begin{equation*}\label{eq:25}
 \eta_2 = 1 - \frac{\sum_{i=1}^{2}  b_i \times f_i }{64 \times rc} - \frac{7}{64} 
\end{equation*}

\begin{equation*}\label{eq:26}
 \eta_2 = 1 - \frac{  1 \times \frac{rc}{2} + 64 \times \frac{rc}{2} }{64 \times rc} - \frac{7}{64} 
\end{equation*}

\begin{equation*}\label{eq:26}
 \eta_2 = 1 - \frac{  32.5  \times rc }{64 \times rc} - \frac{7}{64} 
\end{equation*}

\begin{equation*}\label{eq:27}
 \eta_2 \approx 1 - 0.5078 - 0.1093
\end{equation*}

\begin{equation*}\label{eq:28}
 \eta_2 \approx 38.29\%
\end{equation*}

The efficiency of the VLB method is influenced by the relative size of the bit-length groups. In this first example we considered only two groups, each comprised of half the matrix elements. Let's now vary the relative frequency of the groups, $\frac{f_i}{rc}$, while sticking to two groups. Let's also assume that $\frac{f_i}{rc}$ is a good approximation to the probability of a given bit-length in a matrix, which we will denote by $p_i$.

With this definition we can rewrite the Equation \ref{eq:17}, which becomes \ref{eq:29}. In Equation \ref{eq:29}, the $\frac{f_i}{rc}$ is replaced by $p_i$, representing the probability of elements from group $i$ in matrix M.

\begin{equation}\label{eq:29}
 \eta_2 = 1 - \frac{\sum_{i=1}^{g}  b_i \times p_i }{64} - \frac{k}{64} 
\end{equation}

\noindent with

\begin{equation}\label{eq:30}
 p_i = \frac{f_i}{rc}
\end{equation}

With the Equation \ref{eq:29} can analyze the influence of bit-length probability in compression efficiency. In this example, $p_1$ and $p_2$ represent the probability of elements of bit-lengths 1 and 64, respectively. Thus, efficiency is defined in Equation \ref{eq:31}.

\begin{equation}\label{eq:31}
 \eta_2 = 1 - \frac{1 \times p_1  + 64 \times p_2}{64} - \frac{7}{64} 
\end{equation}

Now, we can determine which probabilities give us the best and worst compression levels. When $\eta_2=1$, then the efficiency is maximal and if $\eta_2=0$, a efficiency is minimal. To calculate the values ​​of $p_1$ and $p_2$ for both extreme values of $\eta_2$,  we must solve the linear systems shown in Equations \ref{eq:32} and \ref{eq:33}. The first equation on both systems come from the law of total probability. The second comes from \ref{eq:31} after setting $\eta_2$ to $1$ and $0$, respectively.

\begin{equation}\label{eq:32}
  \eta_2 = 1 :\left
  \{\begin{matrix}
    p_1 + p_2 = 1\\ 
    p_1+64p_2 = 7
  \end{matrix}
  \right.
\end{equation}

Solving the system above, we find that  when $p_1=0.9047$ and $p_2=0.0953$, efficiency is maximal, and in this particular case is equal to 87.5\%. 

\begin{equation}\label{eq:33}
  \eta_2 = 0 :\left
  \{\begin{matrix}
    p_1 + p_2 = 1\\ 
    p_1+64p_2 = 57
  \end{matrix}
  \right.
\end{equation}

Thus, when $p_1=0.1111$ and $p_2=0.8889$ the efficiency is minimal for the VLB method. For other combinations see Table \ref{tab:02}. Looking at this table, one can see two negative efficiencies, when ($p_1$,$p_2$) assume the values ​​(0,1) and (0.1,0.9). This correspond th cases when the method increases the memory requirements instead of decreasing it.

\begin{table}[h]
 \centering
 \caption{Combinations $p_1$ and $p_2$ to calculate the efficiency.}
 \begin{tabular}{ccc}
  \hline 
  $p_1$  & $p_2$ & $\eta_2$ \\
  \hline
  0.0	&1.0    &-0.109 \\
  0.1	&0.9	&-0.010 \\
  0.2	&0.8	&0.087 \\
  0.3	&0.7	&0.185 \\
  0.4	&0.6	&0.284 \\
  0.5	&0.5	&0.382 \\
  0.6	&0.4	&0.481 \\
  0.7	&0.3	&0.579 \\ 
  0.8	&0.2	&0.678 \\
  0.9	&0.1	&0.776 \\ 
  1.0	&0.0	&0.875 \\
  \hline
 \end{tabular}
 \label{tab:02}
\end{table}

So far, we have examined only two groups (hence two probabilities) of bit-length for the sake of simplicity. Before we generalize to probability distributions let's take a quick look at the efficiencies for more groups, with uniform probability:

\begin{itemize}
  \item 3 groups with bit-lengths 1, 32 and 64 bits, efficiency $\eta_2=0.3854$,
  \item 5 groups with bit-lengths 1, 16, 32, 48 and 64 bits, efficiency $\eta_2=0.3875$,
  \item 8 groups with bit-lengths 1, 8, 16, 24, 32, 40, 48, 56 and 64 bits, efficiency $\eta_2=0.3888$
\end{itemize}

When the distribution of the group probabilities is uniform, i.e., the groups have approximately the same size, efficiency is basically the same, regardless of the number of groups.

Now we can leverage the notion of bit-length probabilities, and study efficiency when bit-lengths follow some  commonly used discrete probability distributions: Discrete Uniform, Binomial and Poisson. For all the experiments, we assume $k=7$, that is, the maximum possible bit-length is 64 bits. Thus, efficiency obtained will not be the best possible, since for that we would need assume small values of $k$ (Equation \ref{eq:29}). 

\subsection*{Discrete Uniform}
Let the bit-lengths of the matrices be distribute according to the Uniform distribution $U(a=1,b=64)$, which means bit-lengths may take values in the set $\{1, 2, 3, $\ldots$, 64\}$ with equal probability, i.e., $\frac{1}{64}$. 

\paragraph{Theoretical Efficiency:}
Let the random variable $B \sim U(a=1,b=64)$ represent the  bit-length of the elements of matrix $M$. Then $E(b_i) = \sum_{i} b_i \times p(b_i) = \frac{a+b}{2}$.  Applying this result to the expected compression efficiency of VLB method (Equation \ref{eq:29}), we have 

\begin{equation}\label{eq:44}
 E(\eta_2) = 1 - \frac{E(B)}{64} - \frac{k}{64} 
\end{equation}

assuming all bit-lengths are possible, i.e., $a=1$ and $b=64$, and hence $k=7$, we can calculate $\eta_2$:

\begin{equation}\label{eq:45}
 E(\eta_2) = 1 - \frac{\frac{1+64}{2}}{64} - \frac{7}{64} \approx 38.28\% 
\end{equation}

This result agrees with the numerical estimates presented in Table \ref{tab:03}.

\paragraph{Numerical Estimates:}
To calculate the VLB efficiency, we generated a matrices with 100 ($M_{10 \times 10}$), 10,000 ($M_{100 \times 100}$) and 1,000,000 ($M_{1,000 \times 1,000}$) elements with 1, 8, 16, 32 and 64 number of bitss.  The average efficiency (Table \ref{tab:03}) is calculated from a 1,000 replicates of each matrix size. As expected the compression effiency gets better with lower expected bit-length.

\begin{table}[h]
  \centering
  \caption{Compression efficiency of VLB method of samples with bit-lengths coming from a Discrete Uniform distribution $U(a=1,b=64)$. Average efficiency ($\overline{\eta_2}\pm\textrm{SD}$) were calculated over a 1,000 replicates.}
 \begin{tabular}{cccc}
    \hline
    & &Matrix sizes& \\
    \hline
    & &Sample size & \\
    Expected bit-length & 100 & 10,000 & 1,000,000 \\
    \hline
     1&0.8750$\pm$0.0000& 0.8750$\pm$0.0000&0.8750$\pm$0.0000\\ 
     8&0.8202$\pm$0.0031& 0.8203$\pm$0.0003&0.8203$\pm$0.0000\\ 
     16&0.7580$\pm$0.0066& 0.7578$\pm$0.0007&0.7578$\pm$0.0001\\ 
     32&0.6330$\pm$0.0142& 0.6329$\pm$0.0014&0.6328$\pm$0.0001\\ 
     64&0.3826$\pm$0.0288& 0.3828$\pm$0.0028&0.3828$\pm$0.0003\\ 
    \hline
 \end{tabular}
 \label{tab:03}
\end{table}

\subsection*{Binomial Distribution}

For the binomial distribution, we will use $Bin(n, p)$, with the number of trials $n$ representing the greatest possible bit-length in the matrix, and $np$ giving us the expected bit-length.

\paragraph{Theoretical Efficiency:}
Let bit-length $(B)$ be a random variable with Binomial distribution, $B \sim B(n=64,p=0.5)$, $E(b_i) = \sum_{i} b_i \times p(b_i) = n \times p$ and the eficiency becomes (with $k=7$):

\begin{align}\label{eq:47}
 E(\eta_2) &= 1 - \frac{64 \times p }{64} - \frac{k}{64}  \\
  &= 1 - \frac{64 \times 0.5 }{64} - \frac{7}{64} \approx 39.05\% \nonumber
\end{align}

Which again agrees with estimates in Table \ref{tab:04}.

\paragraph{Numerical Estimates:}
For these experiments, the parameter n represents the maximum bit-length of matrix elements and takes values in $\{1, 8, 16, 32, 64\}$. In this case, we evaluate the efficiency as a function of the parameter $n$, and matrix size. Even though efficiency does not depend on matrix size, we tried different sizes to test the stability of the compression algorithm. Results are shown in Table \ref{tab:04}. As expected, smaller bit-lengths lead to higher compression efficiencies.

\begin{table}[h]
  \centering
  \caption{Compression efficiency with bit-lengths distributed according to a binomial distribution $B(n,0.5)$. Parameter $n \in \{1, 8, 16, 32, 64 \} $ represents the maximum bit-length. Since $p=0.5$ the expected bit-length is $n/2$ (first column)}
   \begin{tabular}{cccc}
     \hline
     & &Efficiency& \\
     \hline
     & &Sample size& \\
     Expected bit-length $(np)$ &100 &1,000 &1,000,000 \\
     \hline
     1 &0.8828$\pm$0.0004&0.8828$\pm$0.0004&0.8828$\pm$0.0000\\ 
     8 &0.8283$\pm$0.0022&0.8281$\pm$0.0002&0.8281$\pm$0.0000\\ 
     16&0.7656$\pm$0.0032&0.7656$\pm$0.0003&0.7656$\pm$0.0000\\ 
     32&0.6406$\pm$0.0045&0.6406$\pm$0.0004&0.6406$\pm$0.0000\\ 
     64&0.3910$\pm$0.0065&0.3906$\pm$0.0006&0.3906$\pm$0.0001\\
     \hline
  \end{tabular}
  \label{tab:04}
\end{table}

\subsection*{Poisson Distribution}
With bit-length derived from a Poisson($\lambda$), the parameter $\lambda$ corresponds to the expected bit-length. For the purpose of this analysis this Poisson distribution is truncated at 64. 
\paragraph{Theoretical Efficiency:}
Let bit-length $B \sim Poisson(\lambda=32)$. In this case, $E(b_i) = \lambda$, with $k=7$,  the efficiency becomes:

\begin{align}
 E(\eta_2) &= 1 - \frac{\lambda}{64} - \frac{k}{64} \label{eq:49} \\
 &= 1 - \frac{32}{64} - \frac{7}{64} \approx 39.06\% \label{eq:50}
\end{align}

This result is in accordance to Table \ref{tab:05}.
\paragraph{Numerical Estimates:}
The results for this simulation can be seen in \ref{tab:05}. Note that increasing the number of bits to represent numbers increases, there is a loss of efficiency in the compression process. In this case we did not simulate for $\lambda=64$ since a large portion of the samples would fall above the maximum bit-length we are considering for this analysis. 

 \begin{table}[h]
   \centering
   \caption{Compression efficiency with bit-lengths distributed according to a Poisson distribution ($\lambda$), where $\lambda$ represents the expected bit-length.}
  \begin{tabular}{cccc}
      \hline
      &&Efficiency $\eta_2$      & \\
      \hline
      &&Matrix Size& \\
      Expected bit-length ($\lambda$)	& 100	& 1,000		    & 1,000,000 \\
      \hline
      1 	& 0.8751$\pm$0.0015 	& 0.8750$\pm$0.0004 & 0.8750$\pm$0.0000 \\ 
      8 	& 0.7654$\pm$0.0046 	& 0.7656$\pm$0.0005 & 0.7656$\pm$0.0000 \\ 
      16 	& 0.6405$\pm$0.0064 	& 0.6406$\pm$0.0006 & 0.6406$\pm$0.0001 \\ 
      32 	& 0.3908$\pm$0.0087 	& 0.3906$\pm$0.0009 & 0.3906$\pm$0.0001 \\ 
      \hline
  \end{tabular}
  \label{tab:05}
 \end{table}

These results show that a good compression is guaranteed when bit-lengths are distributed according to the tested distributions regardless of sample size.

\section{Conclusion}
In this paper, we have focused in the compression of matrix data, since this is one of the most important application the authors foresee. However, the compression methodology presented can be applied to any numerical data structure, with gains to performance and memory footprint[citar tese Crysttian e possíveis artigos derivados]. 

Further discussions about doing computation with such compressed data-structures will be the subject of another manuscript (in preparation) in which we will present details about the implementation of the compression algorithm, and benchmarks on classical linear algebra tasks such as those in Linpack[ref].

For the compression calculations presented in this paper we limited bit-lenght of integers to 64 bits. However the compression would work in the same way as discussed for computer architectures with larger word sizes.

Representation of floating point numbers is also possible within the proposed compression framework, but at the expense of precision in their representation. Although this may sound like a limitation, when we take into consideration that most experimental data have fewer ``significant'' digits than the maximal precision available in modern computers, fairly good compression may still be achievable for floats.

% You may title this section "Methods" or "Models". 
% "Models" is not a valid title for PLoS ONE authors. However, PLoS ONE
% authors may use "Analysis" 

% Do NOT remove this, even if you are not including acknowledgments
\section*{Acknowledgments}
We would like to thank Claudia Torres Code\c{c}o, Paulo Cezar Carvalho and Moacyr Silva for fruitful discussions and key ideas which helped improve the manuscript.

%\section*{References}
% The bibtex filename
\bibliography{plos.bib}

%\section*{Figure Legends}
%\begin{figure}[!ht]
%\begin{center}
%%\includegraphics[width=4in]{figure_name.2.eps}
%\end{center}
%\caption{
%{\bf Bold the first sentence.}  Rest of figure 2  caption.  Caption 
%should be left justified, as specified by the options to the caption 
%package.
%}
%\label{Figure_label}
%\end{figure}

%\section*{Tables}
%\begin{table}[!ht]
%\caption{
%\bf{Table title}}
%\begin{tabular}{|c|c|c|}
%table information
%\end{tabular}
%\begin{flushleft}Table caption
%\end{flushleft}
%\label{tab:label}
% \end{table}

\end{document}